\documentclass[10pt,twocolumns]{article}
\usepackage{newlfont}
\usepackage[figtopcap]{subfigure}
\usepackage{amsmath}
\usepackage{times}
\usepackage[all]{xypic}
\usepackage{array,multirow,graphicx}
\usepackage[colorlinks,citecolor=blue,urlcolor=blue,linkcolor=blue]{hyperref}
\usepackage{authblk}
\RequirePackage[numbers,sort&compress]{natbib}
\numberwithin{equation}{section}
\begin{document}
\title{Graceful Exit Inflation in $f(T)$ Gravity}%
\author[1,2,3]{G.G.L. Nashed \thanks{nashed@bue.edu.eg}}
\author[1,3]{W. El Hanafy \thanks{waleed.elhanafy@bue.edu.eg}}
\author[1,2]{Sh.Kh. Ibrahim \thanks{shymaa{\_}77@sci.asu.edu.eg}}
\affil[1]{\small \it Centre for theoretical physics at the British University in Egypt, 11837 - P.O. Box 43, Egypt.}
\affil[2]{\small \it Mathematics Department, Faculty of Science, Ain Shams University, Cairo, Egypt.}
\affil[3]{\small \it Egyptian Relativity Group, Egypt.}
\renewcommand\Authands{ and }
\date{}
\maketitle
\begin{abstract}
We apply a quadratic teleparallel torsion scalar of the $f(T)=T+\alpha T^{2}$ field equations to the spatially flat Friedmann-Robertson-Walker (FRW) model. We assume two perfect fluid components, the matter component has a fixed equation of state (EoS) parameter $\omega$, while the torsion component has a dynamical EoS. We obtain an effective scale factor allowing a graceful exit inflation model with no need to slow roll technique. We perform a standard cosmological study to examine the cosmic evolution. In addition, the effective EoS shows consistent results confirming a smooth phase transition from inflation to radiation dominant universe. We consider the case when the torsion is made of a scalar field. This treatment enables us to induce a scalar field sensitive to the spacetime symmetry with an effective potential constructed from the quadratic $f(T)$ gravity. The model is parameterized by two parameters ($\alpha,\omega$) both derive the universe to exit out of de Sitter expansion. The first is purely gravitational and works effectively at large Hubble regime of the early stage allowing a slow roll potential. The second parameter $\omega$ is a thermal-like correction coupled to the kinetic term and works effectively at low Hubble regime of late stages. The slow roll analysis of the obtained potential can perform tensor-to-scalar ratio and spectral index parameters consistent with the recent Planck and BICEP2 data. Both cosmological and scalar field analyses show consistent results.
\end{abstract}
\section{Introduction}\label{S1}
The formulation of the general relativity (GR) theory is powered by the torsion-less Levi-Civita connection within the Riemannian geometry ($R \neq 0$, $T=0$), where $R$ is the curvature which gives the attraction gravity only. Nevertheless, other geometries with different qualities may give the repulsive side of gravity \cite{W2012,AP2013}. It is known that Levi-Civita connection plays the role of the displacement field in the GR, so we expect different qualities when using the curvature-less Weitzenb\"ock connection of the teleparallel geometry ($R=0$, $T\neq 0$). However, using the later geometry provides a teleparallel gravity theory equivalent to the general relativity (TEGR).

The use of this geometry has been started by Einstein in order to unify gravity and electromagnetism \cite{E28}. Although, this trial did not succeed, the geometrical structure has been developed later \cite{M52, MM56}. Even the first trials of using this geometry to obtain a gauge field theory of gravity have shown a great interest \cite{U56,K61,S64,U80,H76}. Extensions to Einstein's work allowed a class of theories to appear independently taking the Lagrangian density to be quadratic in the torsion tensor \cite{MW77,M78,HS79}. Also, it worths to mention the developments of the teleparallel geometry attempting the global approach using arbitrary moving frames instead of the local expressions in the natural basis \cite{NS07,NS13}. Interestingly, some recent promising developments to the teleparallel geometry have been done to install some Finslerian properties to this geometry \cite{W09, WK11,YST12}.

A new class of modified gravity theories have been developed by attempting an arbitrary function $f(R)$ of the Ricci scalar in the Einstein-Hilbert action instead of the Ricci scalar. Similarly, a different class of modified gravity theories which have recently received an attention are the $f(T)$ gravity \cite{BF09, L10,1008.4036,1011.0508}. The applications of the $f(T)$ in cosmology show an interesting results. For example, avoiding the big bang singularity by presenting a bouncing solution \cite{CCDDS11,CQSW14}. Also, $f(T)$ cosmology provides an alternative tool to study inflationary models \cite{FF07,FF08,BF09,FF11,BNO14,BO14,JMM14,HLOS14,NH14,WH14,HN14,HN214}. Moreover, the problems in $f(T)$ theories are discussed \cite{OINC13,IGO14}.

We next highlight some useful relations between Riemannian and teleparallel geometries. The double contraction of the first Bianchi identity of the teleparallel geometry gives
\begin{equation}
\nonumber    R=-T-2\nabla_{\alpha}T^{\nu \alpha}{_{\nu}},
\end{equation}
where the covariant derivative $\nabla$ is with respect to the Levi-Civita connection. The second term in the right hand side is a total derivative, so it has no contribution in the variation when use the right hand side instead of the Ricci scalar in the Einstein-Hilbert action. So the Ricci and teleparallel torsion scalars are equivalent up to a total derivative term. In spite of this quantitative equivalence they are qualitatively very different. For example, the Ricci scalar is invariant under local Lorentz transformation while the total derivative term is not, so the torsion scalar. Accordingly, the $f(T)$ theories are not invariant under local Lorentz transformation \cite{1010.1041,1012.4039}. On the other hand, it is will known that $f(R)$ theories are conformally equivalent to Einstein-Hilbert action plus a scalar field. In contrast, the $f(T)$ theory cannot be conformally equivalent to TEGR plus scalar field \cite{Y2011}.

Remarkably the $R^{2}$ correction into the $f(R)$ gravity induces an early time inflation, namely the Starobinsky inflation, which is in agreement with Planck data. As mentioned above, the analogue in the $f(T)$ with $T^{2}$ correction term is not expected to give the same inflationary model of $R^{2}$. In addition, the Born-Infeld type theories in Weitzenb\"ock geometry suggested a Lagrangian density \cite{F13,S2014}
\begin{equation}
\nonumber    \mathcal{L}_{BI}=\lambda\left[\left(1+2/d~\frac{T}{\lambda}\right)^{d/2}-1\right],
\end{equation}
where $d$ is the dimension and $\lambda$ is a parameter taken as $\lambda \rightarrow \infty$ for low energy limit. Generally, the Born-Infeld versions of gravity theories are useful to regulate singularities present in Einstein's theory. It seems that the quadratic $f(T)$ is the natural choice of the $4$-dimensional teleparallel equivalent of Born-Infeld type theories. According to the above discussion we mainly study the quadratic $f(T)=T+\alpha T^{2}$ on the flat FRW model, where the quadratic contribution is expected to be much important at early universe. Also, to examine possible cosmological constraints on the value of the coefficient $\alpha$. Actually, a similar study to examine the quadratic effect ($\alpha T^{2}$) has been performed within the solar system regime \cite{IS12}.

The work is organized as follows: In Section \ref{S2}, we briefly review the teleparallel geometry and the $f(T)$ gravity. In Section \ref{S3}, we apply the quadratic $f(T)$ on the spatially flat FRW model with an isotropic perfect fluid. In Section \ref{S4}, we show that the quadratic torsion scalar contribution provides a graceful exit out of inflation to radiation dominant universe with no need to slow roll approximation. In Section \ref{S5}, we perform a cosmological study of the obtained model. In Section \ref{S6}, we consider the case when the torsion potential is made of a scalar field to study the consequences on the early cosmic evolution, this technique allows to map the torsion contribution in the Friedmann equations to a scalar field minimally coupled to gravity. We show that model is parameterized by the quadratic coupling constant $\alpha$ which works effectively at early stages allowing reheating to occur through slow roll potential, which provides a graceful exit out of de Sitter expansion to radiation dominant universe of the standard FRW cosmology. The second is the EoS parameter $\omega$ which works effectively through kinetic term at later stages providing a thermal-like correction, which allows a transition from  radiation to cold dark matter by changing the thermal EoS. The quadratic contribution of the torsion scalar in the $f(T)$ and the scalar field frameworks are equivalent. In Section \ref{S7}, we show that the model capable to predict a range of values of the tensor-to-scalar ratio parameter for the same spectral index parameter corresponds the change of the EoS parameter. This has been discussed in the light of the Planck satellite and BICEP2 data. The work is summarized in Section \ref{S8}.
\section{$f(T)$ Gravity Theories}\label{S2}
\subsection{Teleparallel spacetime}\label{S1.1}
This space is described as a pair $(M,~h_{i})$, where $M$ is an $N$-dimensional smooth manifold and $h_{i}$ ($i=1,\cdots, N$) are $N$ independent vector fields defined globally on $M$. The vector fields $h_{i}$ are called the parallelization vector fields. In the four dimensional manifold the parallelization vector fields are called the tetrad field. Where the covariant derivative of the parallelization vector field should vanish
\begin{equation}\label{q1}
D_{\nu} {h_i}^\mu:=\partial_{\nu}
{h_i}^\mu+{\Gamma^\mu}_{\lambda \nu} {h_i}^\lambda=0,
\end{equation}
where the $D$ operator is with respect to the Weitzenb\"{o}ck connection and $\partial_{\nu}:=\frac{\partial}{\partial x^{\nu}}$ and ${\Gamma^\mu}_{\lambda \nu}$ define the nonsymmetric affine connection \cite{Wr}.
\begin{equation}\label{q2}
{\Gamma^\lambda}_{\mu \nu} := {h_i}^\lambda~ \partial_\nu h^{i}{_{\mu}}.
\end{equation}
We assume a four dimensional manifold, the metric tensor $g_{\mu \nu}$ is defined by
\begin{equation}\label{q3}
 g_{\mu \nu} :=  \eta_{i j} {h^i}_\mu {h^j}_\nu,
\end{equation}
where $\eta_{i j}=(+,-,-,-)$ is the metric of Minkowski spacetime. It can be shown that the metricity condition is fulfilled as a consequence of equation (\ref{q1}). Interestingly, the connection (\ref{q2}) has an identically vanishing curvature tensor $R$ but a non vanishing torsion tensor $T$. While the vanishing of the torsion tensor implies the space to be Minkowiskian. We note that, the tetrad field ${h_i}^\mu$ determines a unique metric $g_{\mu \nu}$, while the inverse is incorrect. The torsion $T$ and the contortion $K$ tensor fields are
\begin{eqnarray}
\nonumber {T^\alpha}_{\mu \nu}  & := &
{\Gamma^\alpha}_{\nu \mu}-{\Gamma^\alpha}_{\mu \nu} ={h_i}^\alpha
\left(\partial_\mu{h^i}_\nu-\partial_\nu{h^i}_\mu\right),\\
{K^{\mu \nu}}_\alpha  & := &
-\frac{1}{2}\left({T^{\mu \nu}}_\alpha-{T^{\nu
\mu}}_\alpha-{T_\alpha}^{\mu \nu}\right). \label{q4}
\end{eqnarray}
We introduce the teleparallel torsion scalar which reproduces the teleparallel equivalent to general relativity (TEGR) theory as
\begin{equation}\label{Tor_sc}
T := {T^\alpha}_{\mu \nu} {S_\alpha}^{\mu \nu},
\end{equation}
where the tensor $S$ of type $(2,1)$ is defined as
\begin{equation}\label{q5}
{S_\alpha}^{\mu \nu} := \frac{1}{2}\left({K^{\mu\nu}}_\alpha+\delta^\mu_\alpha{T^{\beta
\nu}}_\beta-\delta^\nu_\alpha{T^{\beta \mu}}_\beta\right),
\end{equation}
which is skew symmetric in the last two indices.

Dealing with the connection coefficients as displacement fields in the $f(T)$ theories might lead to new physical insight of these theories. We use (\ref{q4}) to reexpress the Weitzenb\"{o}ck connection (\ref{q2}) as
\begin{equation}\label{contortion}
    {\Gamma^\mu}_{\nu \rho }=\left \{_{\nu  \rho}^\mu\right\}+{K^{\mu}}_{\nu \rho}.
\end{equation}
The first is the Levi-Civita connection of the GR theory, this connection is defined by the metric $g_{\mu \nu}$ (gravitational potential) and its first derivatives with respect to the coordinates. This connection contributes as the attractive gravity in the GR. While the second term is made of the contortion (torsion) which consists of the tetrad vector fields ${h_i}^\mu$ and its first derivatives with respect to the coordinates. Similarly, we can think of the second term as a force of gravity. Actually its contribution to the modified geodesics indicates a repulsive gravity \cite{W2012}. The value added quality of the teleparallel space is its capability to describe two faces of gravity, attractive and repulsive. So torsion gravity can provide a unique source to explain early and late phases of the cosmic accelerating expansion \cite{NH14}.
\subsection{$f(T)$ field equations}\label{S1.2}
Similar to the $f(R)$ theory one can defines the action of $f(T)$ theory as
\begin{equation}\label{q7}
S=\frac{M_{\textmd{\tiny Pl}}^2}{2}\int |h|f(T)~d^{4}x+\int {\cal L}_{Matter}({h^i}_\mu,\Phi_A)~d^{4}x,
\end{equation}
where $M_{\textmd{\tiny Pl}}$ is the reduced Planck mass, which is related to the gravitational constant $G$ by $M_{\textmd{\tiny Pl}}=\sqrt{\hbar c/8\pi G}$. Assuming the units in which $G = c = \hbar = 1$, in the above equation $ |h|=\sqrt{-g}=\det\left({h^a}_\mu\right)$, $\Phi_A$ are the matter fields. The variation of (\ref{q7}) with respect to the field ${h^i}_\mu$ requires the following field equations \cite{BF09}
\begin{eqnarray}\label{q8}
\nonumber &{S_\mu}^{\rho \nu} \partial_{\rho} T f_{TT}+\left[h^{-1}{h^i}_\mu\partial_\rho\left(h{h_i}^\alpha
{S_\alpha}^{\rho \nu}\right)-{T^\alpha}_{\lambda \mu}{S_\alpha}^{\nu \lambda}\right]f_T&\\
&-\frac{1}{4}\delta^\nu_\mu f=-4\pi{{\cal T}_{\mu}}^{\nu},&
\end{eqnarray}
where $f := f(T)$, $\;f_{T}:=\frac{\partial f(T)}{\partial T}$, $\;f_{TT}:=\frac{\partial^2 f(T)}{\partial T^2}$.
\section{$f(T)$ Cosmological modifications}\label{S3}
We apply the $f(T)$ field equations (\ref{q8}) to the FRW universe of a spatially homogeneous and isotropic spacetime, which directly gives rise to the tetrad given by Robertson \cite{R32}. This can be written in spherical polar coordinate ($t$, $r$, $\theta$, $\phi$) as follows:
\begin{small}
\begin{eqnarray}\label{tetrad}
\nonumber \left({h_{i}}^{\mu}\right)=
\left(
  \begin{array}{cccc}
    1 & 0 & 0 & 0 \\
    0&\frac{L_1 \sin{\theta} \cos{\phi}}{4a(t)} & \frac{L_2 \cos{\theta} \cos{\phi}-4r\sqrt{k}\sin{\phi}}{4 r a(t)} & -\frac{L_2 \sin{\phi}+4 r \sqrt{k} \cos{\theta} \cos{\phi}}{4 r a(t)\sin{\theta}} \\[5pt]
    0&\frac{L_1 \sin{\theta} \sin{\phi}}{4 a(t)} & \frac{L_2 \cos{\theta} \sin{\phi}+4 r \sqrt{k}\cos{\phi}}{4 r a(t)} & \frac{L_2 \cos{\phi}-4 r \sqrt{k} \cos{\theta} \sin{\phi}}{4 r a(t)\sin{\theta}} \\[5pt]
    0&\frac{L_1 \cos{\theta}}{4 a(t)} & \frac{-L_2 \sin{\theta}}{4 r a(t)} & \frac{\sqrt{k}}{a(t)} \\[5pt]
  \end{array}
\right),\\
\end{eqnarray}
\end{small}
where $L_1=4+k r^{2}$ and $L_2=4-k r^{2}$ and $a(t)$ is the scale factor. The EoS is taken for an isotropic fluid so that the energy-momentum tensor is ${{\cal T}_{\mu}}^{\nu}=\textmd{diag}(\rho,-p,-p,-p)$. The tetrad (\ref{tetrad}) has the same metric as FRW metric. Substituting from (\ref{tetrad}) into (\ref{Tor_sc}) we evaluate the torsion scalar as
\begin{equation}\label{Tsc}
    T=-6H^{2}(1+\Omega_{k}),
\end{equation}
where $H(t):=\frac{\dot{a}(t)}{a(t)}$ is the Hubble parameter, the dot denotes the derivative with respect to time, and $\Omega_{k}:=\frac{-k}{a^{2} H^{2}}$ is the curvature density parameter. In order to introduce the torsion contribution to the density and pressure in the Friedmann dynamical equations we replace $\rho \rightarrow \rho+\rho_{T}$ and $p \rightarrow p+p_{T}$, thus the $f(T)$ field equations (\ref{q8}) read
\begin{eqnarray}
  3H^2    &=& 8\pi (\rho+\rho_{T}) - 3\frac{k}{a^2},\label{TFRW1} \\
  3 q H^2 &=& -4 \pi \left[(\rho+\rho_{T})+3 (p+p_{T})\right], \label{TFRW2}
\end{eqnarray}
where $\rho$ and $p$ are the total density and pressure of the matter inside the universe
\begin{eqnarray}
\rho&=&\frac{1}{16 \pi}(f+12 H^2 f_{T}),\label{dens1}\\
\nonumber p&=&-\frac{1}{16 \pi}\left[(f+12 H^2 f_{T}) + 4\dot{H}(f_{T}-12 H^2 f_{TT})\right.\\
&-&\left.\frac{4k}{a^2}(f_{T}+12H^2 f_{TT})\right].\label{press1}
\end{eqnarray}
The above equations are the modified Friedmann equations in the FRW universe governed by $f(T)$ gravity. The torsion contributes to the energy density and pressure as \cite{NH14}
\begin{eqnarray}
\rho_{T}&=&\frac{1}{8 \pi}\left(3H^2-f/2-6H^2 f_{T}+\frac{3k}{a^2}\right),\label{Tor_dens}\\
\nonumber p_{T}&=&\frac{-1}{8 \pi}\left[\frac{k}{a^2}(1+2f_{T}+24H^2 f_{TT})+2\dot{H}+3H^2\right.\\
         &-&\left.f/2-2(\dot{H}+3H^2)f_{T}+24\dot{H}H^2 f_{TT}\right].\label{Tor_press}
\end{eqnarray}
The equations (\ref{dens1})-(\ref{Tor_press}) can perform the following cases: (i) The case of ($k=0$, $f(T)=T$), the torsion contribution vanishes while $\rho$ and $p$ reduce to the Friedmann equations of GR of flat space. (ii) The case of ($k\neq 0$, $f(T)=T$), the torsion contributes as in Friedmann equations as anti-curvature density, or simply as a cosmological constant. (iii) The case of ($k=0$, $f(T)\neq T$), it reduces to \cite{M2011}, where the matter and torsion have dynamical contributions. (iv) The case of ($k\neq 0$, $f(T)\neq T$), this case shows the dynamical evolution in non-flat models \cite{NH14}.
\section{The quadratic $f(T)$ gravity theory}\label{S4}
One of the most successful inflationary models in $f(R)$ theories is the $R+R^{2}$ theory. When a conformal transformation on the quadratic $f(R)$ action from Jordan frame to Einstein frame is performed, it can be shown that the theory is equivalent to Einstein-Hilbert action plus a scalar field $\varphi$ with a potential
\begin{equation}
\nonumber    V_{S}(\varphi) \sim \left(1-e^{-\sqrt{2/3}\varphi}\right)^{2}.
\end{equation}
The above expression referred to as Starobinsky potential. The slow roll analysis of this potential predicts a spectral index $n_{s} \sim 0.96$ and tensor-to-scalar ratio $r \sim 4.8 \times 10^{-3}$ when the number of $e$-folds $N=50$. So this inflationary model is in agreement with Planck observations \cite{1303.5076,1303.5082}, with much smaller tensor-to-scalar ratio than the upper limit ($r<0.11$) suggested by Planck. On contrary, the model cannot perform large values of the tensor-to-scalar ratio suggested by BICEP2 ($r \sim 0.2$) \cite{1403.3985}. However, recent work has shown that Starobinsky model having a double attractors. In addition to the success of the model to induce an early-time cosmic inflation, it gives a unified description of the inflation with dark energy and cancelation of finite-time future singularities \cite{BMOS14}. Also, it has been shown that the $R^{2}$ theory cannot be the de Sitter inflation so that the quadratic correction has been studied in the string theory to generate a graceful exit from inflation. Similar analogue in the $f(T)$ theory is not expected. In order to examine the analogue behaviour of the quadratic $f(T)$ theory of gravity, we take the action without matter as below
\begin{equation}\label{fT}
    S=\frac{M_{\textmd{\tiny Pl}}^2}{2} \int |h|\left(T+\alpha T^{2}\right)~d^{4}x,
\end{equation}
where $\alpha$ is a constant parameter. It is clear that this form produces TEGR theory when $\alpha=0$. As the $f(T)$ function in the FRW spacetime is a function of time $f(T \rightarrow t)$, one easily can show that
\begin{equation}\label{Fd2T}
  f_{T} = \dot{f}/\dot{T},\quad  f_{TT} = \left(\dot{T} \ddot{f}-\ddot{T} \dot{f}\right)/\dot{T}^3.
\end{equation}
For the flat space universe ($k=0$) the equations (\ref{dens1})-(\ref{Tor_press}) now read
\begin{eqnarray}
\rho&=&\frac{3}{8\pi}\frac{\dot{a}^{2}(a^{2}-18\alpha\dot{a}^{2})}{a^{4}},\label{dens2}\\
p&=&-\frac{1}{8 \pi}\frac{a^{2}\dot{a}^{2}+2\ddot{a}a^{3}+\alpha(18\dot{a}^{4}-72\dot{a}^{2}\ddot{a}a)}{a^{4}}.\label{press2}
\end{eqnarray}
The matter density and pressure reduces to the GR results when $\alpha=0$. However, the torsion contributes to the total density and pressure as
\begin{eqnarray}
\rho_{T}&=&\frac{27\alpha}{4 \pi}\frac{\dot{a}^{4}}{a^{4}},\label{Tor_dens2}\\
p_{T}&=&-\frac{9\alpha}{4 \pi}\frac{\dot{a}^{2}(4\ddot{a}a-\dot{a}^{2})}{a^{4}}.\label{Tor_press2}
\end{eqnarray}
The torsion contribution vanishes where $\alpha=0$, so the theory generally produces the GR theory.
\subsection{Eternal inflation}\label{S4.1}
Substituting from (\ref{dens2}) and (\ref{press2}) into the matter continuity equation
\begin{equation}\label{cont1}
    \dot{\rho}+3H(\rho+p)=0.
\end{equation}
We study the case when the universe is dominated by a cosmological constant dark energy ($\Lambda$DE) like, i.e. the EoS parameter $\omega=-1$. Solving for the scale factor $a(t)$ we get
\begin{equation}\label{sc_fac_vac}
    a_{\textmd{inf}}(t)=a_{0}e^{H_{0}(t-t_{0})},
\end{equation}
where $a_{0}$ is an arbitrary constant of integration, with an initial condition $a_{0} = a(t_{0})$. Consequently, we can evaluate
\begin{equation}\label{THW1}
    T=-\frac{1}{6\alpha},~H=\frac{1}{6\sqrt{\alpha}},~q=-1,~\omega_{T}=-1.
\end{equation}
The solution gives an exponential expanding universe with a constant Hubble parameter, i.e. de Sitter universe. So equivalently we have a vacuum dominant universe which powered the inflation scenario. Recalling (\ref{fT}), we find the coefficient $\alpha$ of the $T^{2}$ higher order gravity contributes to this inflationary phase. The comparison of the results (\ref{THW1}) with the de Sitter solution directly relates the coefficient $\alpha$ to the cosmological constant $\Lambda$ by $\alpha = \frac{1}{12 \Lambda}$. It is well known that at this early stage the cosmological constant has a value much larger than its present value. The discrepancy between the two values is about 120 orders of magnitude, this is called the cosmological constant problem. According to our analysis; we expect the value of $\alpha$ is very small $\sim 10^{-73}~s^{2}$, so that the quadratic contribution in the $f(T)$ can be ignored, which is unlikely at early universe. We next examine the second-fluid (torsion) component. Substituting from (\ref{sc_fac_vac}) into (\ref{Tor_dens2}) and (\ref{Tor_press2}) we find that the torsion fluid satisfies the continuity equation as
\begin{equation}\label{cont2}
    \dot{\rho}_{T}+3H(\rho_{T}+p_{T}) \equiv 0,
\end{equation}
so we find that the total energy density satisfies the continuity equation. This case shows that both the matter and the torsion fluids having the same behaviour where $\rho=\rho_{T}=\frac{1}{192\pi \alpha}=\frac{\Lambda}{16\pi}$, which is the density of vacuum. Also, the EoS parameters has been obtained as $\omega_{T}:=p_{T}/\rho_{T}=-1$. This case gives directly a de Sitter universe, where both matter and torsion act as a vacuum state in an indistinguishable behaviour. Using (\ref{fT}) and (\ref{THW1}) we get $f(T)=-\frac{5}{36\alpha}\sim T$. Interestingly, we find that the solution omits the second order contribution of the $f(T)$ naturally and the $f(T)$ is a constant, more precisely it acts just like the cosmological constant. This may explain why the two-fluid components have the same behaviour. We saw that the choice of $\omega=-1$ of the matter fluid constrains the torsion fluid to acquire the same behaviour producing a pure vacuum universe which represents the key of the inflationary universe. Although the model does not address the cosmological constant issue, it succeeded to predict the exponential cosmic inflation. But it does not provide a mechanism to end the inflation epoch and resuming the big bang to radiation dominant universe era.

In conclusion, the $\omega=-1$ model produces a de Sitter expansion case of the universe. These types of inflationary models are completely invariant under space and time translations so that they are incapable of ending the inflation epoch. Although, the de Sitter inflation is a useful approximation at late time, the early universe inflation cannot be exactly de Sitter.
\subsection{Graceful exit inflation without slow-roll}\label{S4.2}
Similarly, we take the second case when the EoS parameter of the matter fluid $\omega \neq -1$. The series solution of the continuity equation of matter (\ref{cont1}) up to $O(\alpha^{2})$ provides the following scale factor
\begin{eqnarray}\label{sc_fac_matt}
    a_{\textmd{eff}}(t)&=&a_{0}\underbrace{\left[(1+\omega)(t-t_{0})\right]^{\frac{2}{3(1+\omega)}}}_{\textmd{matter}}\underbrace{e^{\frac{-8 \alpha}{(1+\omega)^{3}(t-t_{0})^{2}}}}_{\textmd{vacuum}},\\
    &=&a_{0}.~ a_{m}(t).~ a_{\textmd{v}}(t),
\end{eqnarray}
where $a_{0}$ and $t_{0}:=\tau_{0}-\frac{4(2+27\alpha H_{0}^{2})}{9H_{0}(1+\omega)}+O(\alpha^2)$ are arbitrary constants of integration, with an initial condition $H_{0} = H(\tau_{0})$. The effective scale factor (\ref{sc_fac_matt}) can be decomposed into two terms, where the term $\propto t^{\frac{2}{3(1+\omega)}}$ represents the different matter dominant epochs, it takes the values of $\sim t^{1/2}$ and $\sim t^{2/3}$ for radiation ($\omega=1/3$) and dust ($\omega=0$) states, respectively. This perfectly matches the TEGR theory. While the quadratic $T^{2}$ correction contributes in (\ref{sc_fac_matt}) as an exponential function of time representing an inflationary (vacuum dominant) epoch. So we take $a_{m}$ and $a_{\textmd{v}}$ in (\ref{sc_fac_matt}) to denote the scale factor of matter and vacuum dominant epochs, respectively. In order to describe how the universe evolves according to the scale factor (\ref{sc_fac_matt}), we concisely determine the behaviour of the scale factor (\ref{sc_fac_matt}) when $t \rightarrow t_{0}$ and its asymptotic behaviour when $t \rightarrow \infty$. So we get
\begin{equation}
\nonumber    \lim_{t\rightarrow t_{0}}a_{m}=0,~\lim_{t\rightarrow \infty}a_{\textmd{v}}=1.
\end{equation}
The above limits show that the scale factor is dominated by $a_{\textmd{v}}$ where $a_{m}$ is negligible at the very early universe time; then the $a_{\textmd{v}}$ approaches its saturation level just after a sharp growth, while the effect of $a_{m}$ becomes dominant at later phase. This is in agreement with \cite{FF07} where the scale factor is initially exponential. In this model, we investigate beyond the initial exponential which leads to late decelerated FRW of the standard cosmology. So we say that the scale factor starts with an inflationary universe with an exponential scale factor decays smoothly to new phase of radiation dominant universe.
\begin{figure}
\centering
\subfigure[figtopcap][phase diagram]{\label{fig1a}\includegraphics[scale=.3]{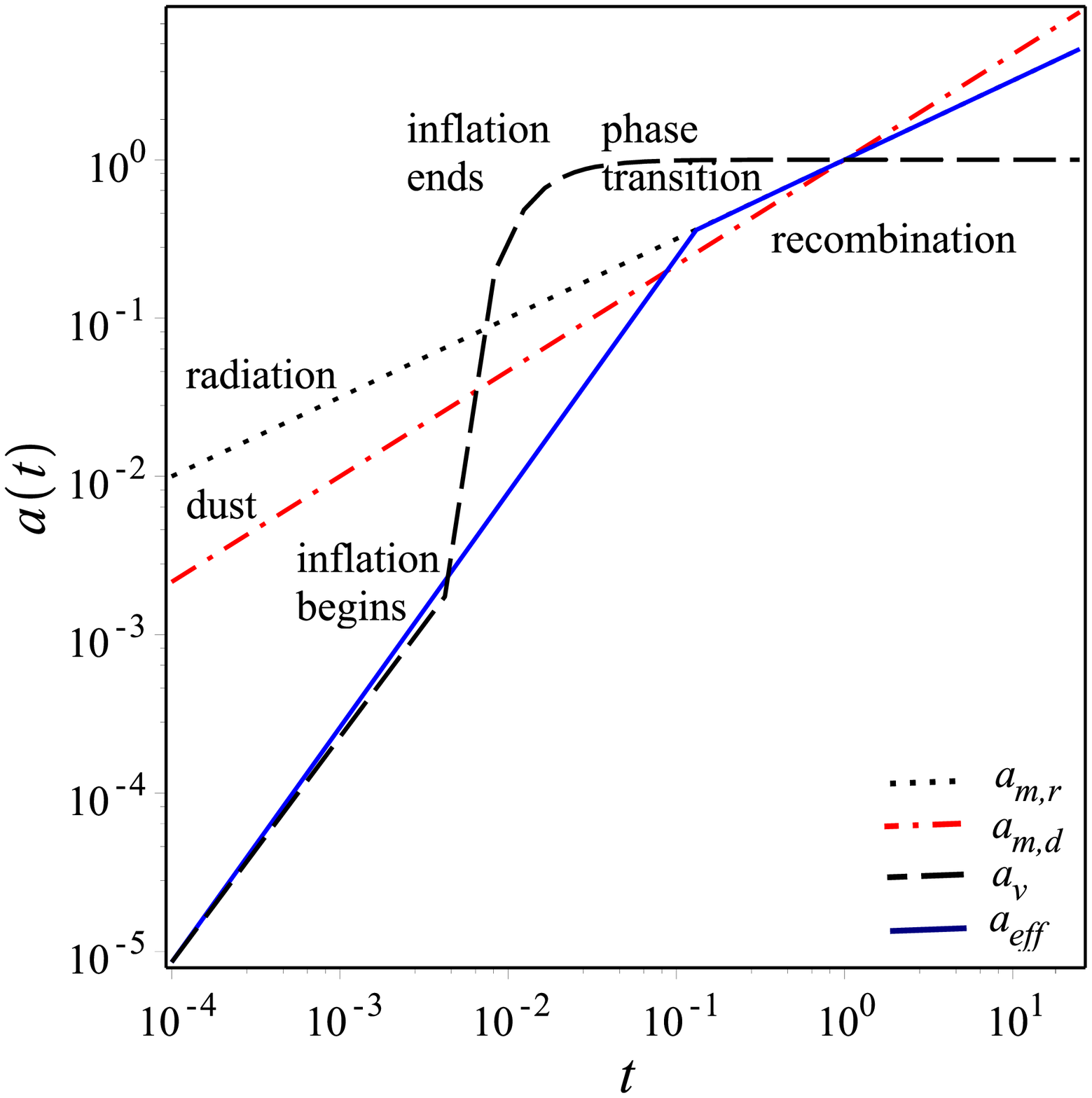}}\hspace{2cm}
\subfigure[figtopcap][graceful exit inflation]{\label{fig1b}\includegraphics[scale=.3]{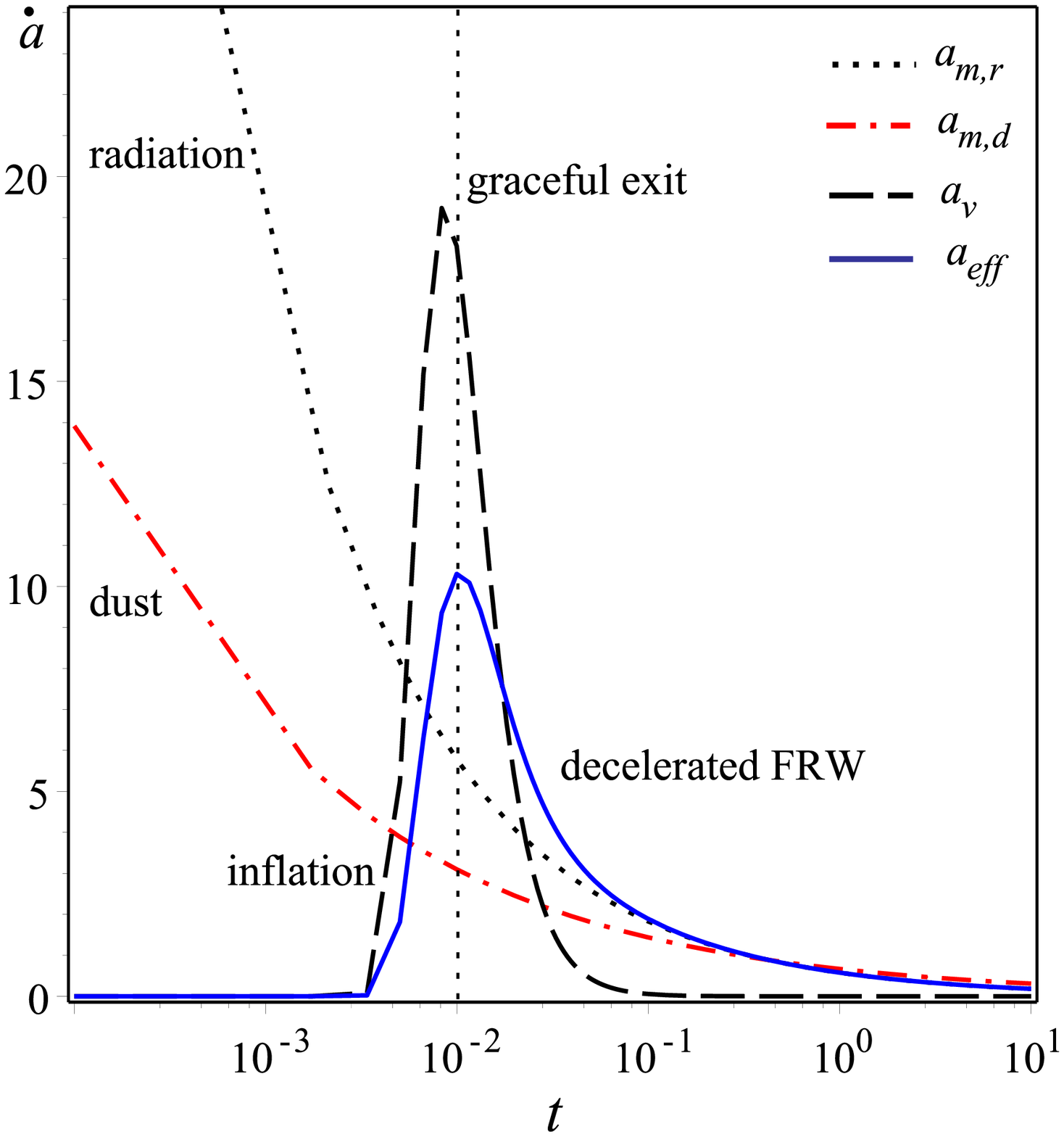}}
\caption[figtopcap]{Phase transitions according to the scale factor (\ref{sc_fac_matt}):
\subref{fig1a} the plots show the evolution of the scale factor vs. cosmic time, where $a_{m,r}$ and $a_{m,d}$ denote the radiation ($\omega=1/3$) and dust ($\omega=0$), respectively;
\subref{fig1b} the rate of change of the effective scale factor plot shows a typical pattern of a graceful exit inflation model to the standard FRW cosmology. The initial values have been chosen as $t_{0}=10^{-5}$ and $a_{0}=1$. The model parameter is chosen as $\alpha=10^{-4}$. The constant $a_{0}$ controls the amplitude of $a_{\textmd{eff}}$, where the parameter $\alpha$ determines the time of the inflation.}
\label{Fig1}
\end{figure}
The plots in Figure \ref{Fig1}\subref{fig1a} show the universe phase diagram of its different ingredients of matter and vacuum states providing a clear early phase transition from an inflation epoch to radiation dominant universe. Also, the plots in Figure \ref{Fig1}\subref{fig1b} show the rate of change evolution of the scale factor, the plot provides a typical pattern of graceful inflation models. From the phase transition plot we see that three different stages summarized as follows:

(\textbf{i}) The first stage at the early time shows that both matter and vacuum having similar behaviour. Although the effective scale factor (\ref{sc_fac_matt}) matches the steeper evolution of the vacuum state, so the vacuum dominates radiation at this pre-stage.

(\textbf{ii}) The second stage shows that the vacuum evolves much faster than radiation matter providing an inflationary phase; then the vacuum shows a steady expansion ending the inflation period, while the radiation evolves similar to the first stage. By the end of this intermediate stage the two components follow different tracks. The radiation energy decreases as the frequency $\nu \propto a(t)^{-1}$.
    \begin{equation}
    \nonumber    \frac{a_{\textmd{end}}}{a_{\textmd{inf}}}=\frac{\nu_{\textmd{inf}}}{\nu_{\textmd{end}}}=\frac{E_{\textmd{inf}}}{E_{\textmd{end}}}.
    \end{equation}
    where $E$ is the energy of the radiation, the subscripts ``inf" and ``end" denote the beginning and the end of the second stage (inflation phase), respectively. Nevertheless, the vacuum approaches a critical value $a_{\textmd{v}} \rightarrow 1$ whereas the vacuum has no more repulsive energy to loose during the supercooling process of the inflation. Now the vacuum switch the inflation off as given by the plots of Figure \ref{Fig1}\subref{fig1a}.

(\textbf{iii}) In the last stage the radiation evolves much faster than vacuum. While the flat plateau of the vacuum in Figure \ref{Fig1} indicates that the variation of its scale factor vanishes, which leads to a vanishing of an energy transfer (by cooling) associated to expansion. This is similar to the case of the cooling curve when a substance enters a phase transition. This stage describes a vacuum decaying process in which the vacuum releases its latent energy allowing a new phase of reheating the universe producing a relativistic radiation dominant epoch. This is consistent with the late behaviour of the effective scale factor that matches perfectly the ultra-relativistic (radiation) dominant, till the recombination at the radiation-dust equality (i.e. $a_{m,r}=a_{m,d}$) where the universe is cooled enough to turn on the non-relativistic (dust) matter.\\
It is well known that the inflation has provided an add-on tool to solve the big bang cosmology problems. In quantum cosmology, the inflationary universe is fairly understood using the scalar field description. In the old inflation model the graceful exit mechanism of inflation from false to true vacuum is trough quantum tunneling event, while the new inflation model provides an alternative mechanism which requires a scalar field with a very flat plateau at the false vacuum slowly rolls to its effective minimum at the true vacuum. This mechanism is called slow roll mechanism. However, the Born-Infeld-TEGR gravity can provide an alternative tool to inflation without inflaton scalar field \cite{FF07}. Similar strategy has been applied in $f(T)$ gravity \cite{F13}. However, these treatments strongly succeeded to replace the initial spacetime singularity by a de Sitter expansion, it did not provide a mechanism to exit out of de Sitter expansion. Actually, a recent study in this context of Born-Infeld-$f(T)$ gravity related the obtained models to late expansion rather the early one \cite{S2014}. In this model, we show that the quadratic $T^{2}$ contribution provides a graceful exit inflation model without slow roll approximation. However, this model provides a satisfactory mechanism to enforce the vacuum to end its inflation and enters a phase transition state releasing its latent energy to allow the radiation to dominate the universe. Also, Figure \ref{Fig1}\subref{fig1b}, at the large Hubble regime the rate of change of the effective scale factor $\dot{a}_{\textmd{eff}}$ is dominated by the exponential expansion where the coupling to matter is negligible so the plot shows an early accelerated phase (inflation) behaviour $\dot{a}_{\textmd{eff}}$ increases. In contrast, at small Hubble regime where the matter contribution becomes more effective it derives the inflation to deviate from de Sitter expansion to late deceleration of the standard FRW cosmology where the phase transition now is powered by the change of the thermal EoS parameter, this is shown in Figure \ref{Fig1}\subref{fig1b} as the $\dot{a}_{\textmd{eff}}$ decreases. So the model provides an alternative mechanism to the turn the universe from an inflationary phase to radiation dominant phase naturally without imposing slow-roll conditions. However, we will show later, in Section \ref{S6}, that the scalar field description can be achieved by using the torsion potential treatment.
\section{Cosmological study}\label{S5}
\begin{figure}
\centering
\subfigure[figtopcap][Hubble parameter]{\label{fig2a}\includegraphics[scale=.3]{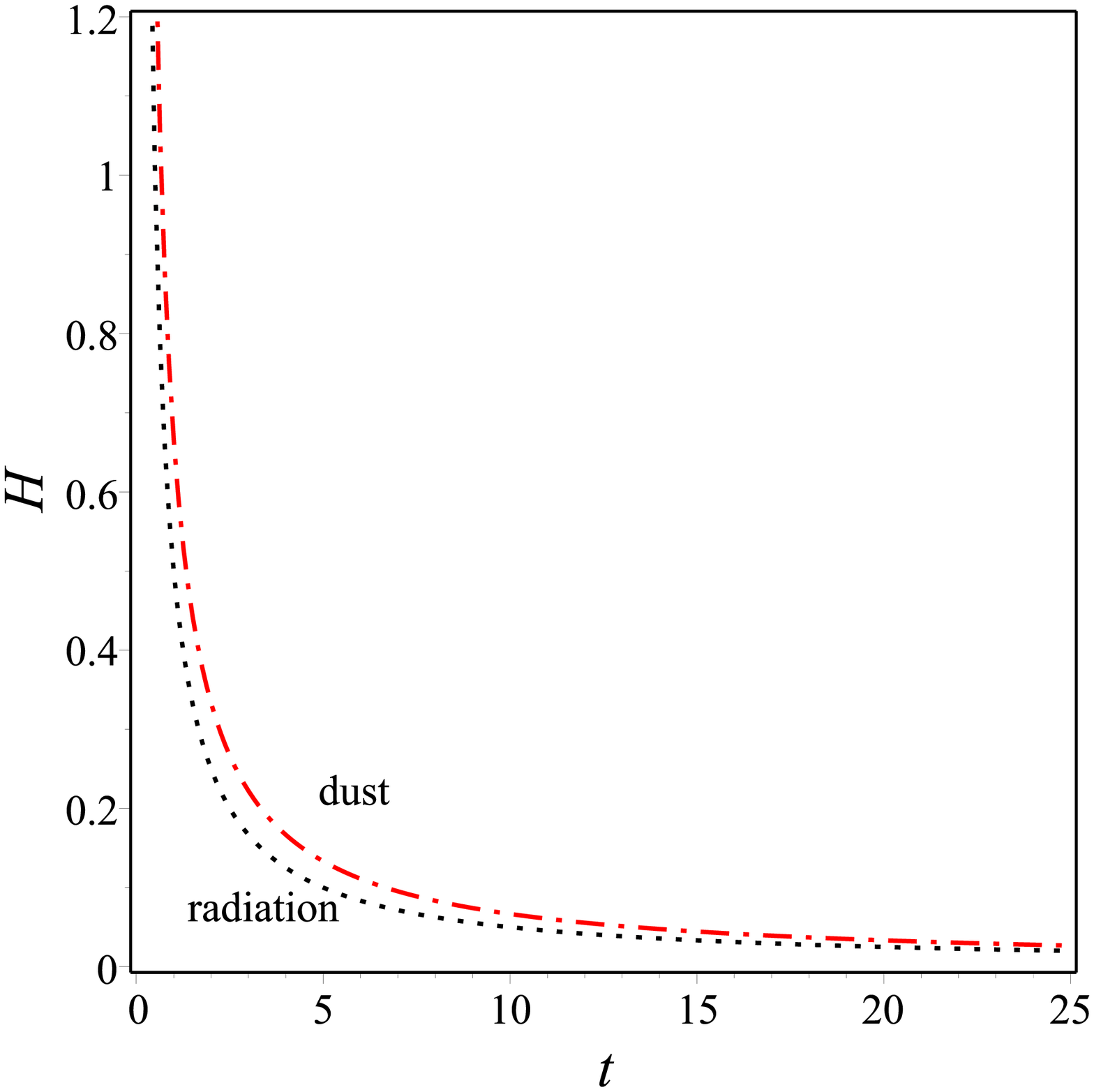}}\hspace{2cm}
\subfigure[figtopcap][deceleration parameter]{\label{fig2b}\includegraphics[scale=.3]{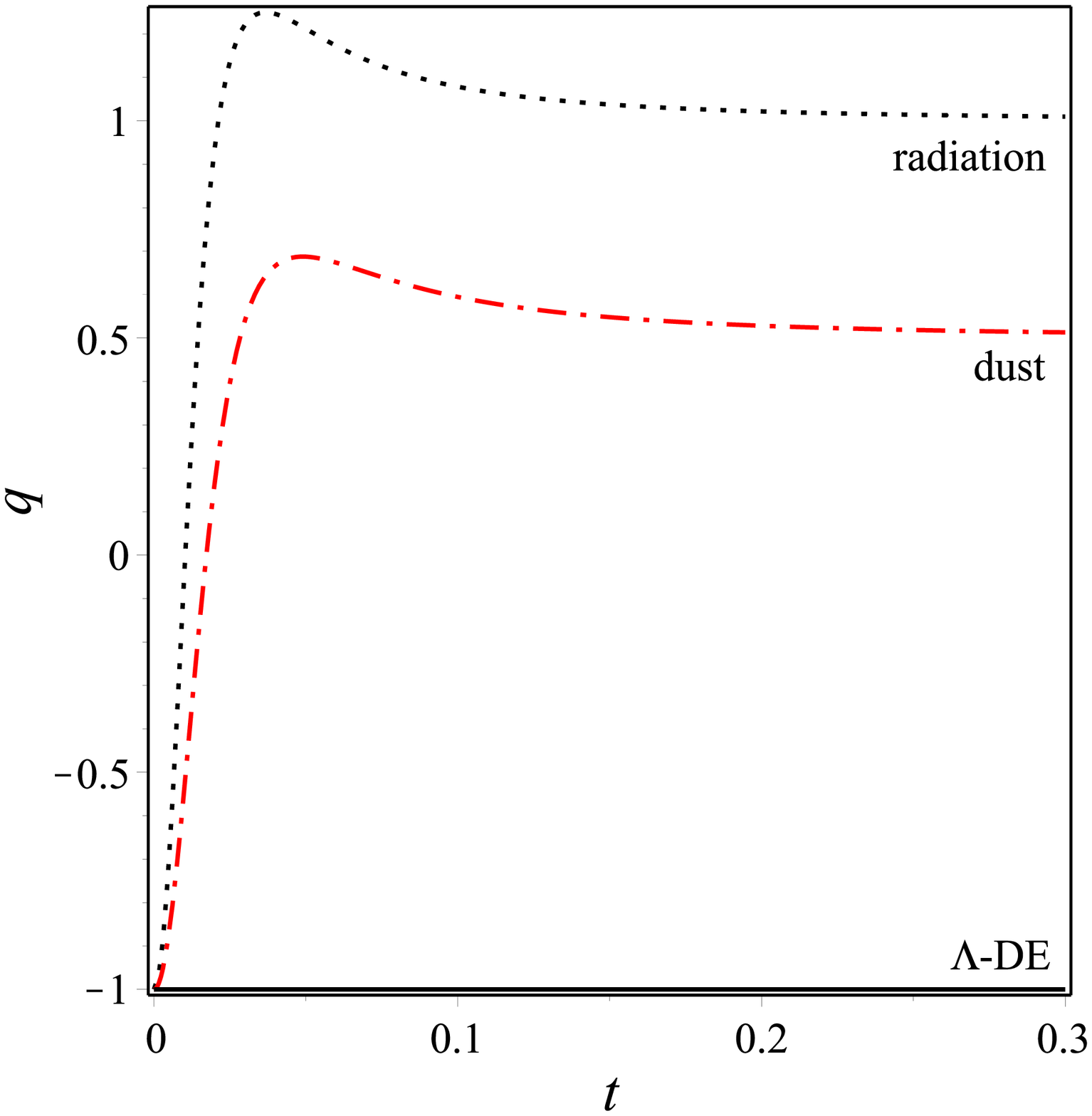}}
\caption[figtopcap]{The cosmological parameters:
\subref{fig1a} the evolution of the Hubble parameter (\ref{Hubb2}) indicates an early cosmic inflation;
\subref{fig1b} the evolution of the deceleration parameter $q:=-\frac{a\ddot{a}}{\dot{a}^{2}}$ indicates a transition from early accelerating to late decelerating cosmic expansion. The constants $t_{0}$ and $a_{0}$ in addition to the parameter $\alpha$ have the same values as in Figure \ref{Fig1}.}
\label{Fig2}
\end{figure}

\begin{figure}
\centering
\subfigure[figtopcap][torsion decay]{\label{fig3a}\includegraphics[scale=.25]{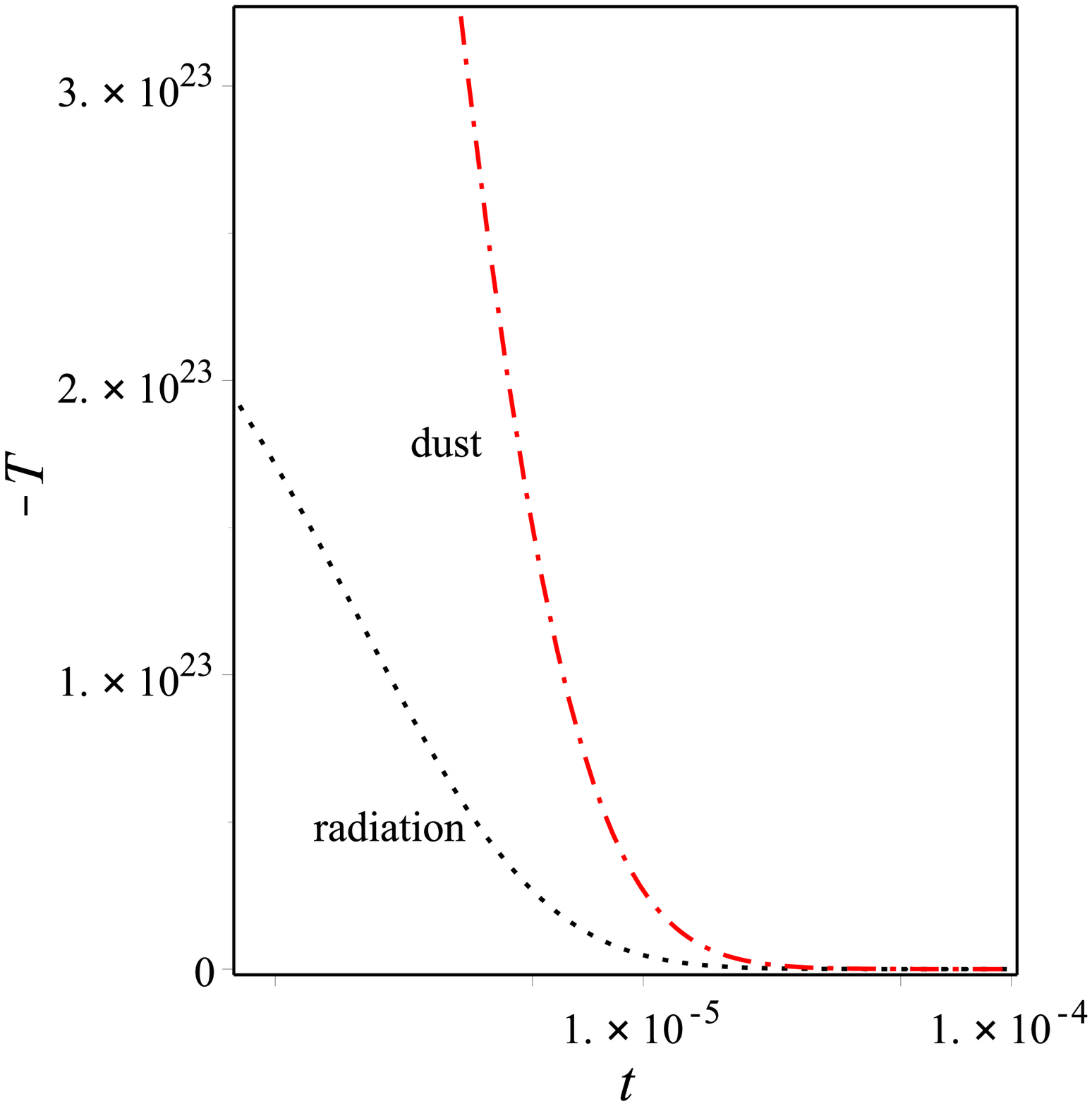}}\hspace{1pt}
\subfigure[figtopcap][torsion EoS]{\label{fig3b}\includegraphics[scale=.25]{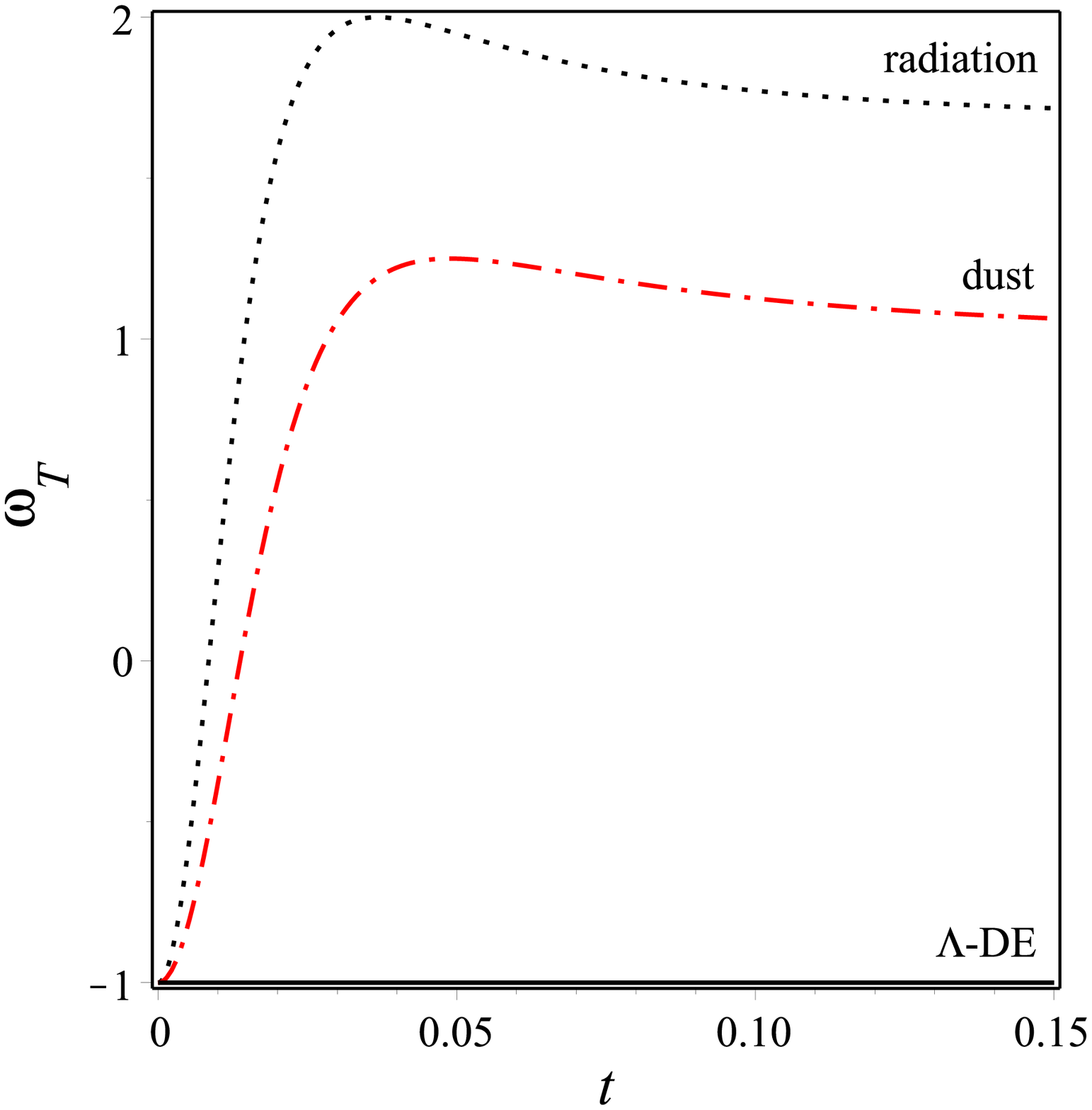}}
\subfigure[figtopcap][effective EoS]{\label{fig3c}\includegraphics[scale=.25]{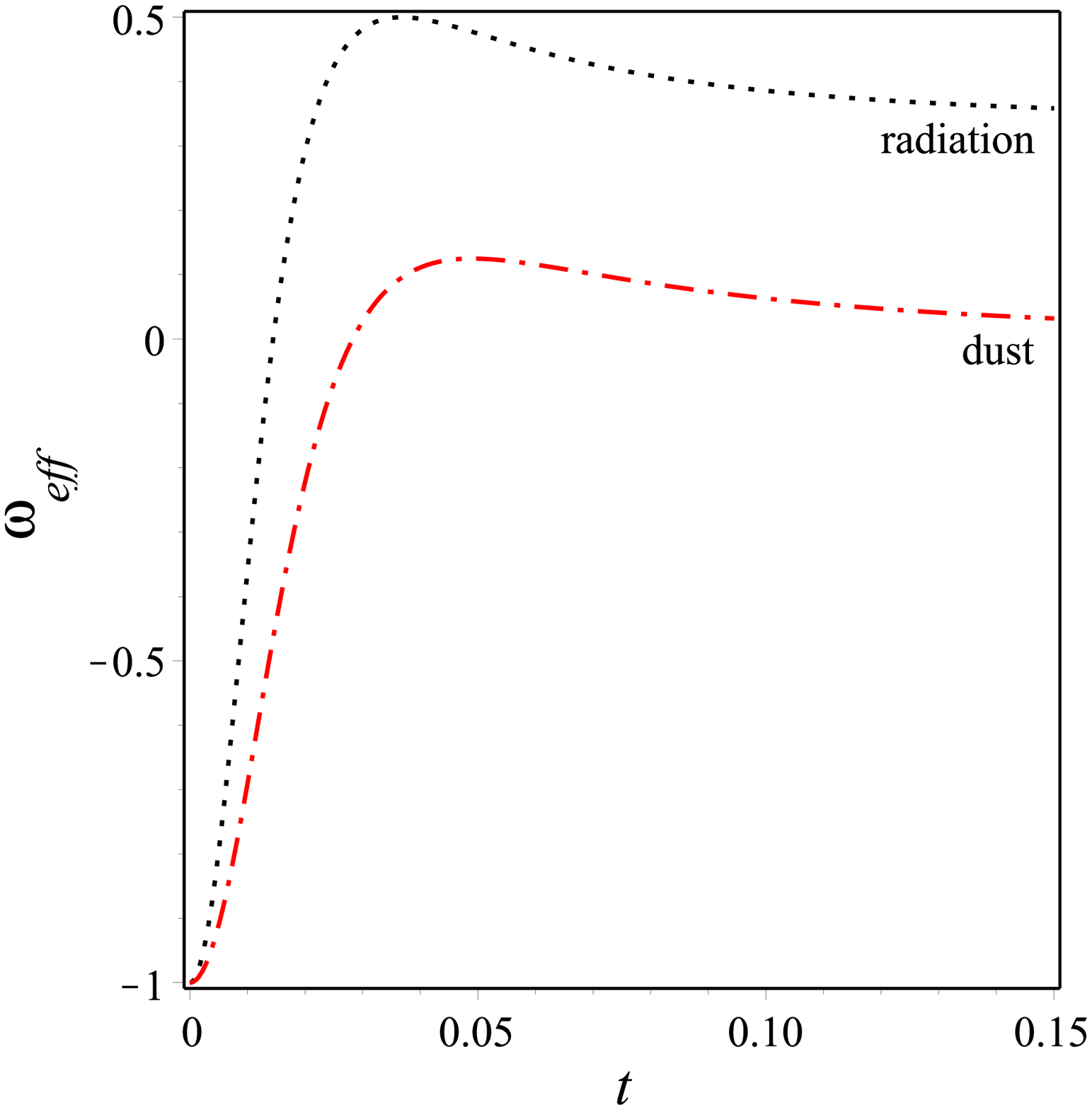}}
\caption[figtopcap]{Cosmic evolution:
\subref{fig3a} the decay of the teleparallel torsion scalar field (\ref{Tsc2});
\subref{fig3b} the evolution of torsion EoS (\ref{Tor_EoS});
\subref{fig3c} the evolution of the effective EoS $\omega_{\textmd{eff}}=(p+p_{T})/(\rho+\rho_{T})$. The constants $t_{0}$ and $a_{0}$ in addition to the parameter $\alpha$ have the same values as in Figure \ref{Fig1}.}
\label{Fig3}
\end{figure}

The standard cosmology gives a set of parameters to study the evolution of the universe through the Hubble and the deceleration parameters, while the other parameters are called density parameters allow studying the composition of the universe. We study some of these parameters correspond to the effective scale factor (\ref{sc_fac_matt}) and the assumed two-fluids which are given by the set of equations (\ref{dens2})-(\ref{Tor_press2}). So the Hubble parameter reads
\begin{equation}\label{Hubb2}
    H(t)=\frac{2(1+\omega)^{2}(t-t_{0})^{2}+16\alpha}{3(1+\omega)^{3}(t-t_{0})^{3}}.
\end{equation}
One can see that when the initial value $t_{0}$ is chosen as the Planck time, the Hubble parameter counts a large value at this early time, while it decays at later time, $\lim_{t\rightarrow \infty}H=0$. Interestingly, the above expression shows a time dependent Hubble parameter capable to describe an inflationary and a later matter dominant phases. Also, the deceleration parameter shows a consistent results with the above description. We evaluate the deceleration parameters at both early and late phases as
\begin{equation}\label{decel}
    \lim_{t\rightarrow t_{0}}q(t)=-1,~\lim_{t\rightarrow \infty}q(t)=\frac{3}{2}\omega+\frac{1}{2}.
\end{equation}
This again indicates an accelerating expansion phase as $q<0$ at the early time, while the universe turns to a decelerating phase at later phase of the matter dominant universe. The asymptotic behaviour is consistent with the known results as $q \rightarrow 1/2~ \textmd{or}~ 1$ when the EoS of the matter are chosen as ($\omega= 0)$ for dust or ($\omega = 1/3$) for radiation, respectively. Also, equation (\ref{decel}) shows that the deceleration $q \rightarrow -1$ at all time when the matter is chosen as $\Lambda$DE ($\omega \rightarrow -1$). This covers the case which has been studied in Subsection (\ref{S4.1}). A more detailed description can be seen from the plots of Figure \ref{Fig2}.
Substituting from (\ref{Hubb2}) into (\ref{Tsc}), noting that $\Omega_{k}=0$ of the flat space, we evaluate the teleparallel torsion scalar as
\begin{equation}\label{Tsc2}
    T(t)=-\frac{8}{3}\frac{\left[(1+\omega)^{2}(t-t_{0})^{2}+8\alpha\right]^{2}}{(1+\omega)^{6}(t-t_{0})^{6}}.
\end{equation}
It is well known that the inflationary models assume a scalar field (inflaton) powering this epoch; then it decays rapidly providing a reheating phase allows the pair production to occur. On the other hand, the torsion tensor field plays the main role in the teleparallel spacetime, i.e. the vanishing of the torsion implies an annihilation of all the teleparallel geometric structure producing a Minkowiskian geometry. Most of the quantum applications require this Minkowiskian background, while one cannot accept this assumption easily at the early time when the matter is condensed in a tiny space producing a highly curved background. We see that the teleparallel geometry, powerfully, gives a description of gravity better and more complete (attractive and repulsive) than the Riemannian's. In addition, the vanishing of the curvature tensor provides a good chance to perform some quantum applications. Interestingly, we show that the teleparallel torsion scalar field of this theory shares the inflaton its rapid decaying character, since
\begin{equation}
\nonumber    \lim_{t\rightarrow t_{0}}T(t)=-\infty,~\lim_{t\rightarrow \infty}T(t)=0.
\end{equation}
A more detailed description can be seen in Figure \ref{Fig3}\subref{fig3a}. The plot shows a rapid decay of the torsion scalar which contributes directly to explain the huge difference between the large cosmological constant at the early universe and its small present value which is known as the cosmological constant problem, for more details see \cite{HN214}. In the next Section, we analog description using scalar field treatment. This leads to study the torsion fluid as a physical vacuum phase, so we evaluate the EoS parameter of the torsion fluid ($\omega_{T}:=p_{T}/\rho_{T}$) at the early and late stages of the universe. Using (\ref{Tor_dens2}) and (\ref{Tor_press2}) we obtain
\begin{equation}\label{Tor_EoS}
    \lim_{t\rightarrow t_{0}}\omega_{T}(t)=-1,~\lim_{t\rightarrow \infty}\omega_{T}(t)=2\omega+1.
\end{equation}
The above evaluation indicates that the torsion fluid initially acts as $\Lambda$DE ($\omega_{T}=-1$) in all cases, crosses the quintessence limit to an ordinary matter with $\omega > 0$ at later stages. We study when the matter is dust and radiation so the asymptotic behaviour of the torsion EoS approaches the values of $\omega_{T} \rightarrow 1$ or $\omega_{T} \rightarrow 5/3$, respectively. Moreover, the equation (\ref{Tor_EoS}) shows that the torsion does not evolve, $\omega_{T} \rightarrow -1$ as $t \rightarrow \infty$, when the second fluid is also $\Lambda$DE. So the torsion may show different evolution according to the type of the associated matter, see Figure \ref{Fig3}\subref{fig3b}, which strongly recommends a certain interaction between the two fluid components of the universe. Finally, we study the effective EoS $\omega_{\textmd{eff}}:=\frac{p+p_{T}}{\rho+\rho_{T}}$ at both early and late time. Using the equations (\ref{dens2})-(\ref{Tor_press2}) we obtain
\begin{eqnarray}
\nonumber    \lim_{t\rightarrow t_{0}}\omega_{\textmd{eff}}(t)=-1,~\lim_{t\rightarrow \infty}\omega_{\textmd{eff}}(t)=\omega,\\
\nonumber\textmd{i.e.}~   \omega_{\textmd{eff}}(t): \omega_{T}(t) \rightarrow \omega \textmd{ as } t:t_{0}\rightarrow \infty.
\end{eqnarray}
The above limits show that the effective EoS initially starts as $\Lambda$DE evolves to match perfectly the matter EoS at later stages. So the universe turns ends its inflationary (torsion dominant) phase, smoothly turns itself to matter dominant phase with no need to slow-roll conditions, Figure \ref{Fig3}\subref{fig3c}. This is consistent with the results of the previous Section.
%
\section{Torsion potential}\label{S6}
We showed that the contortion contributes as a repulsive force of gravity in the modified geodesic equation. This repulsive gravity has the upper hand to power the early or the late cosmic accelerating expansion. On the hand, the scalar field analysis is the most powerful tool to describe the cosmic inflation at the early universe. So we consider the approach that has been purposed by \cite{XSH96}, by introducing sixteen fields $t^{\mu}{_{i}}$ that are called $``$\textit{torsion potential}$"$. These fields form a quadruplet basis vectors, so we write the following linear transformation:
\begin{equation}
\nonumber    h_{i}=t^{\mu}{_{i}}\partial_{\mu},~~h^{i}=t^{i}{_{\mu}}dx^{\mu},
\end{equation}
the torsion potential $t^{\mu}{_{i}}$ and its inverse satisfy
\begin{equation}
\nonumber    t=\textmd{det}(t^{\mu}{_{i}})\neq 0,~~t^{\mu}{_{i}}t^{i}{_{\nu}}=\delta^{\mu}_{\nu},~~t^{\mu}{_{i}}t^{j}{_{\mu}}=\delta^{j}_{i}.
\end{equation}
Then the torsion can be written as \cite{XSH96}
\begin{equation}\label{torsion_pot}
    T^{\alpha}{_{\mu \nu}}=t^{\alpha}{_{i}}\left(\partial_{\nu}t^{i}{_{\mu}}-\partial_{\mu}t^{i}{_{\nu}}\right).
\end{equation}
Generally, the torsion potential $t^{\mu}{_{i}}$ can be reformed by a physical scalar, vector or tensor fields. In the previous sections we showed that the torsion gravity can provide a graceful exit cosmic inflation to radiation dominant universe with no need to the slow roll approximation. However, the cosmic inflation is powered by a spin-$0$ matter, we assume the case when the torsion potential is constructed by a scalar field $\varphi(x)$ with FRW background in order to relate to reformulate the $f(T)$ results using a scalar field, so we take
\begin{equation}
\nonumber    t^{i}{_{\mu}}=\delta^{i}_{\mu}e^{\sqrt{3/2}\varphi},~~t^{\mu}{_{i}}=\delta^{\mu}_{i}e^{-\sqrt{3/2}\varphi},
\end{equation}
where $\varphi$ is a non-vanishing scalar field. Then the torsion is expressed as
\begin{eqnarray}
   T^{\alpha}{_{\mu \nu}}=\sqrt{3/2}\left(\delta^{\alpha}_{\nu}\varphi_{,\mu}-\delta^{\alpha}_{\mu}\varphi_{,\nu}\right),\\
    K^{\mu\nu}{_{\alpha}}=\sqrt{3/2}\left(\delta^{\nu}_{\alpha}\varphi^{,\mu}-\delta^{\mu}_{\alpha}\varphi^{,\nu}\right),\label{semi-symm-torsion}
\end{eqnarray}
where $\varphi^{,\mu}:=g^{\mu \alpha}\varphi_{,\alpha}$. Actually, similar forms of the torsion tensor has been used successfully to make the torsion satisfy the gauge invariance and the minimal coupling principles. This helped to get a dynamical torsion gauge invariant and minimally coupled electromagnetism \cite{R74,HRR78,H90,HO01}. Using the above equations and (\ref{q5}), the teleparallel torsion scalar (\ref{Tor_sc}) can be written in terms of the scalar field $\varphi$ as
\begin{equation}\label{Tsc_phi}
    T=-9\varphi_{,\mu}\varphi^{,\mu}.
\end{equation}
The above treatment shows that the torsion acquires dynamical properties and it  propagates through space. As shown in the previous Section that the teleparallel torsion scalar has a decaying performance while the Weitzenb\"{o}ck connection shows a capability to explain the repulsive gravity. On the other hand, it well known that the cosmic inflation is powered by assuming a spinless particles at this early universe stage. Consequently, we can map the torsion contribution in the Friedmann equations into the scalar field ($\rho_{T}\rightarrow \rho_{\varphi}$, $p_{T}\rightarrow p_{\varphi}$). This leads to reformulate the Friedmann equations of the torsion contribution as an inflationary background in terms of the scalar field $\varphi$. At the inflationary epoch the matter contribution can be negligible, so we consider the Lagrangian density of a homogeneous (real) scalar field $\varphi$ in potential $V(\varphi)$
\begin{equation}\label{lag_dens}
    \mathcal{L}_{\varphi}=\frac{1}{2}\partial_\mu \varphi~ \partial^\mu \varphi-V(\varphi).
\end{equation}
where the first term in the above equation represents the kinetic term of the scalar field, as usual, while $V(\varphi)$ represents the potential of the scalar field. The variation of the action with respect to the metric $g_{\mu \nu}$ enables to define the energy momentum tensor as
\begin{equation}
 \nonumber   \mathcal{T}^{\mu \nu} = \frac{1}{2}\partial^\mu \varphi~ \partial^\nu \varphi-g^{\mu \nu} \mathcal{L}_{\varphi}.
\end{equation}
The variation with respect to the scalar field reads the scalar field density and pressure respectively as
\begin{equation}\label{press_phi1}
    \rho_{\varphi}=\frac{1}{2}\dot{\varphi}^2+V(\varphi),~~ p_{\varphi}=\frac{1}{2}\dot{\varphi}^2-V(\varphi).
\end{equation}
The Friedmann equation (\ref{TFRW1}) in absence of matter becomes
\begin{equation}
\nonumber    H^2=\frac{1}{2}\dot{\varphi}^2+V(\varphi).
\end{equation}
Using (\ref{press_phi1}) it is an easy task to show that continuity equation of the torsion contribution (\ref{cont2}) can be mapped to the Klein-Gordon equation of homogeneous scalar field in the expanding FRW universe
\begin{equation}
\nonumber    \ddot{\varphi}+3H\dot{\varphi}+V'(\varphi)=0,
\end{equation}
where the prime denotes the derivative with respect to the scalar field $\varphi$. In conclusion, equation (\ref{Tsc_phi}) enables to define a scalar field sensitive to the vierbein field, i.e. the spacetime symmetry. In addition, equation (\ref{press_phi1}) enables to evaluate an effective potential from the adopted $f(T)$ gravity theory. The mapping from the torsion contribution to scalar field fulfills the Friedmann and Klein-Gordon equations. So the treatment meets the requirements to reformulate the torsion contribution in terms of a scalar field without using attempting a conformal transformation.
\subsection{Thermal-like correction of the scalar potential}\label{S6.1}
Using equations (\ref{Tsc2}) and (\ref{Tsc_phi}) the temporal gradient of the scalar field is evaluated as
\begin{equation}\label{kin}
\dot{\varphi}=\pm \frac{2\sqrt{6}}{9(1+\omega)}\left[\frac{1}{(t-t_{0})}+\frac{8\alpha}{(1+\omega)^{2}(t-t_{0})^{3}}\right].
\end{equation}
Integrating (\ref{kin}) with respect to time; then the scalar field is given by
\begin{equation}\label{phi}
    \varphi=\varphi_{0} \pm \frac{2\sqrt{6}}{9(1+\omega)}\left[\ln{(t-t_{0})}-\underbrace{\frac{4\alpha}{(1+\omega)^{2}(t-t_{0})^{2}}}_{\textmd{correction term}}\right],
\end{equation}
where $\varphi_{0}$ is a constant of integration. The scalar field $\varphi$ is parameterized by two parameters ($\alpha,\omega$), we first study the effect of the EoS parameter $\omega$ in the scalar field, separately from $\alpha$, by considering the evolution due to the leading term of (\ref{phi}) only such that
\begin{equation}\label{phi_asym}
    \varphi=\varphi_{0} \pm \frac{2\sqrt{6}}{9(1+\omega)}\ln{(t-t_{0})},
\end{equation}
alternatively, we write the parameterized time
\begin{equation}\label{time}
    t=t_{0}+e^{\pm \frac{3\sqrt{6}}{4}(1+\omega)(\varphi-\varphi_{0})}.
\end{equation}
We introduce a new scalar field $\psi:=\frac{3\sqrt{6}(1+\omega)}{4}(\varphi-\varphi_{0})$ to simplify expressions. Equation (\ref{time}) enables us to investigate the transformation between the teleparallel torsion scalar $T$ and the scalar $\psi$ fields. Substituting from (\ref{time}) into (\ref{Tsc2}) so
\begin{equation}
\nonumber    T(\psi)=-\frac{512}{3}e^{\pm 6\psi}\left[\frac{(1+\omega)^{2}e^{\mp 2\psi}+8\alpha}{8(1+\omega)^{3}}\right]^{2},
\end{equation}
\begin{figure}
\centering
\subfigure[figtopcap][the quadratic $\alpha T^{2}$ correction escapes the de Sitter inflation, providing a slow roll inflation model with a pre-tunneling]{\label{fig4a}\includegraphics[scale=.24]{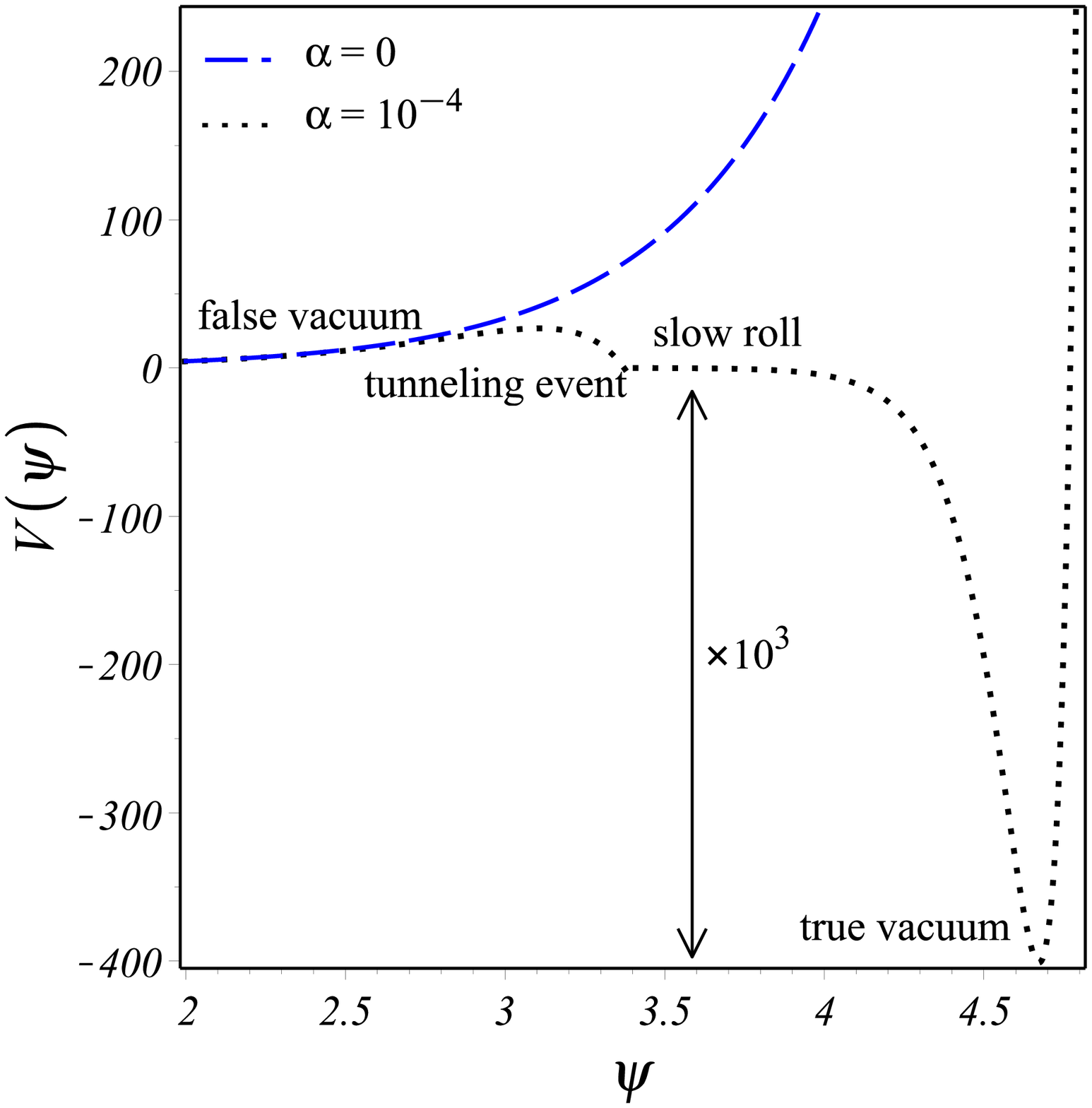}}\hspace{10pt}
\subfigure[figtopcap][the potential pattern at different values of the EoS parameter $\omega$ with a fixed value of the parameter $\alpha=10^{-3}$]{\label{fig4b}\includegraphics[scale=.25]{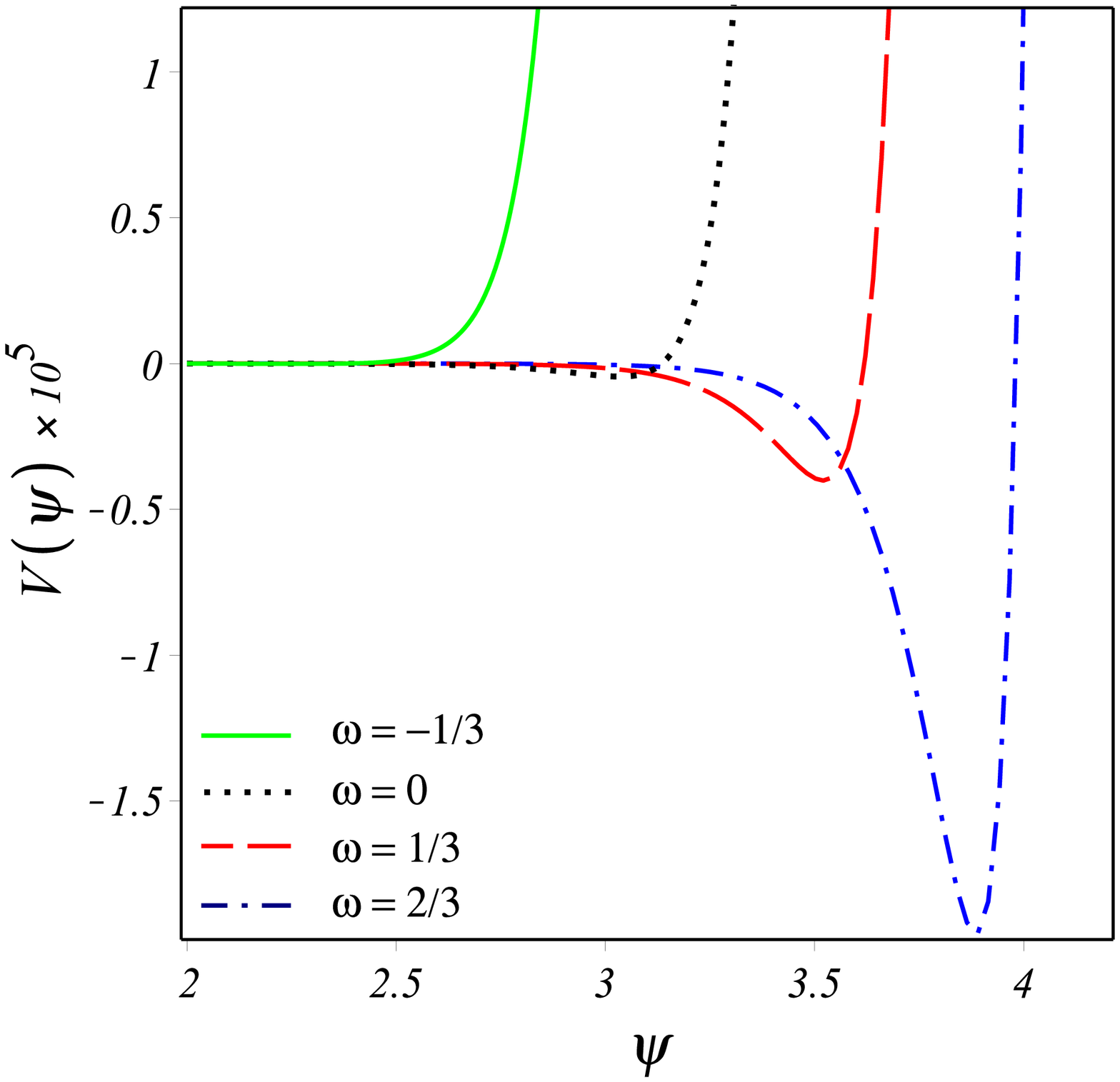}}\hspace{10pt}
\subfigure[figtopcap][the potential pattern at different values of the parameter $\alpha$ with a fixed EoS parameter $\omega=1/3$]{\label{fig4c}\includegraphics[scale=.25]{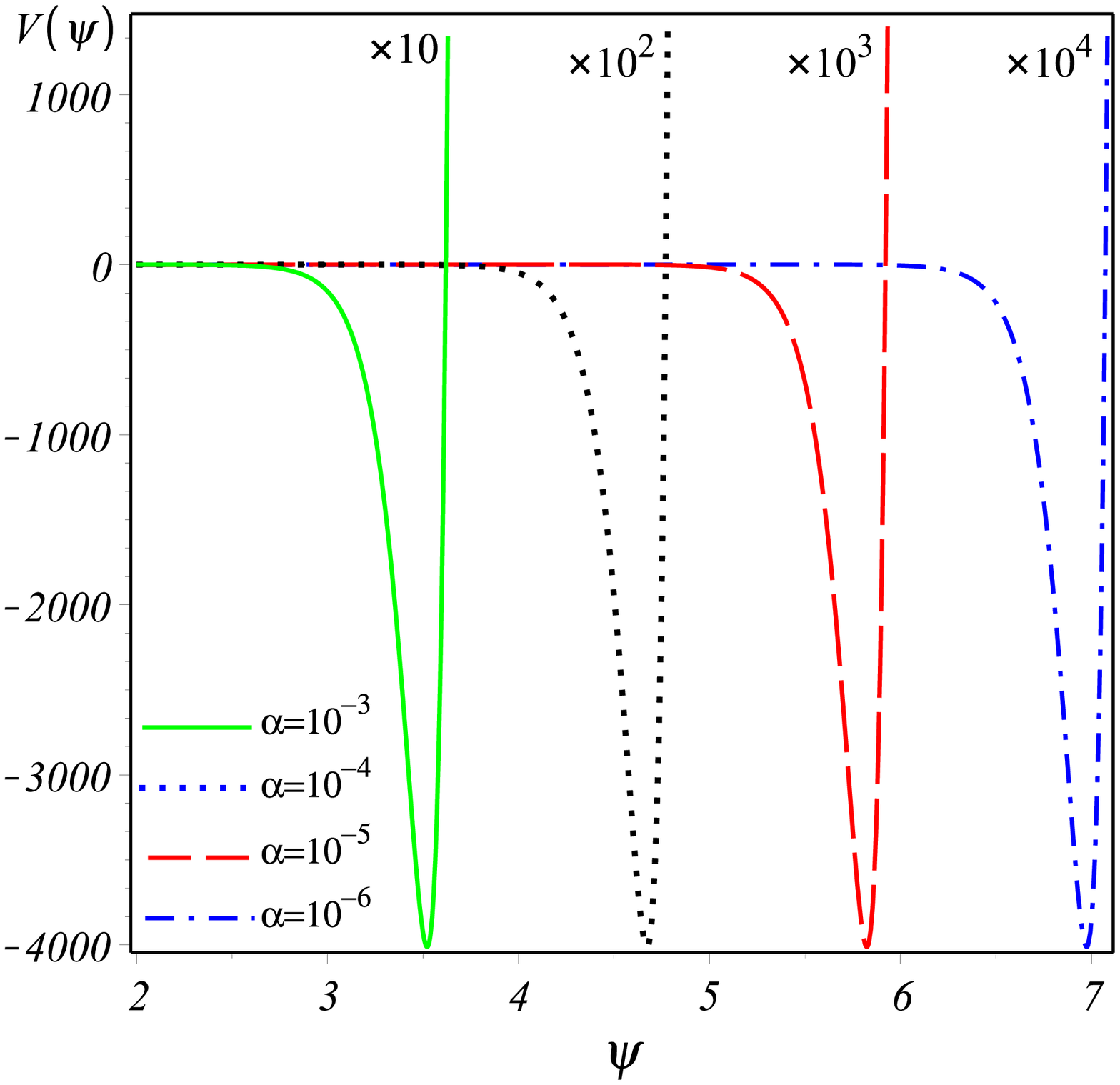}}
\caption[figtopcap]{The potential pattern corresponds to the leading term of the scalar field (\ref{phi_asym}):
\subref{fig4a} the dash line represents eternal $\Lambda$ de Sitter inflation, the dot line shows slow roll potential with a small tunneling around $\psi \sim 3.11$ after false vacuum followed by slowly rolling plateau down to its effective minimum at $\psi \sim 4.67$;
\subref{fig4b} the plots show the EoS correction derives the potential $V(\psi_{\textmd{min}})>0$ as $\omega<0$ so that the evolution cannot escape the de Sitter universe, while $V(\psi_{\textmd{min}})<0$ as $\omega>0$ allowing a slow roll inflation.
\subref{fig4c} the plots show the sensitivity of the potential towards the choice of the parameter $\alpha$, the small tunneling events do not appear due to the large plot scale. The initial values have been chosen as $t_{0}=10^{-5}$, $\varphi_{0}=3$.}
\label{Fig4}
\end{figure}
For simplicity we write all expressions in terms of the scalar field $\psi$. So using (\ref{sc_fac_matt}), (\ref{Tor_press2}) and (\ref{time}), the torsion pressure (\ref{Tor_press2}) can be reexpressed in terms of the scalar field $\psi$ as
\begin{eqnarray}\label{press_phi2}
\nonumber p_{\psi}&=&\frac{8\alpha(\omega+\frac{1}{2})e^{\pm 4\psi}}{3\pi(1+\omega)^{4}}+\frac{320\alpha^{2}(\omega+\frac{3}{5})e^{\pm 6\psi}}{3\pi(1+\omega)^{6}}\\
\nonumber&+&\frac{3584\alpha^{3}(\omega+\frac{4}{7})e^{\pm 8\psi}}{3\pi(1+\omega)^{8}}+\frac{4096\alpha^{4}(\omega+\frac{1}{3})e^{\pm 10\psi}}{\pi(1+\omega)^{10}}\\
&-&\frac{16384\alpha^{5}e^{\pm 12\psi}}{3\pi(1+\omega)^{12}}=-\frac{32}{3 \pi} \sum_{n=2}^{6} \alpha^{n}\beta_{n} e^{\pm 2n \psi},
\end{eqnarray}
where $\beta_{n}$ are known constant coefficients. Substituting from (\ref{kin}), (\ref{time}) and (\ref{press_phi2}) into (\ref{press_phi1}), we evaluate the potential of the leading scalar field $\psi$ as
\begin{equation}\label{sc_pot}
V_{l}(\psi,\alpha,\omega)=\frac{4e^{\pm 2\psi}}{27(1+\omega)^{2}}+\frac{32}{3 \pi} \sum_{n=2}^{6} \alpha^{n} \beta_{n} e^{\pm 2n\psi}.
\end{equation}
It is clear that the potential up to the power series $n=2$ reduces to Starobinsky model, while the extra four terms are corrections. Next, we study the physical meaning of the EoS parameter contribution in the scalar potential. Recalling the EoS of perfect fluids
\begin{equation}
\nonumber    p=\rho(\gamma-1)C_{\textmd{v}}\Theta=\omega \rho,
\end{equation}
where $\Theta$ is the absolute temperature, $\gamma=C_{p}/C_{\textmd{v}}$ is the adiabatic index (ratio of specific heats), $C_{p}$ and $C_{\textmd{v}}$ are the specific heat at constant pressure and volume, respectively. Actually the EoS parameter correction in (\ref{sc_pot}) is suggested to represent a thermal correction of the scalar potential as $\omega$ $\propto$ $\Theta$. This is supported by the plots of Figure \ref{Fig4}\subref{fig4b} where the minimum potential $V_{l}(\psi_{\textmd{min}})<0$ is negative as $\omega\geq 0$. In contrast $V_{l}(\psi_{\textmd{min}})=V(0) \geq 0$ when $\omega<0$ such that the universe cannot escape the de Sitter expansion. On the other hand, the second parameter $\alpha$ contribution is suggested to be purely gravitational as appears in the original $f(T)$ theory (\ref{fT}) as a coupling constant of a self-interacting torsion scalar field. The $\alpha$ dependence terms in (\ref{sc_pot}) leads the early expansion when the quadratic $T^{2}$ contribution is important, the sensitivity of the effective potential to the parameter $\alpha$ is shown in Figure \ref{Fig4}\subref{fig4c}. At strong coupling we can introduce a canonical scalar field $\Omega$ such that $\psi=\pm\log(\Omega)$. Now we rewrite the effective potential (\ref{sc_pot}) as
\begin{equation}\label{lead_pot}
    V_{l}(\Omega,\alpha,\omega) \sim \frac{\Omega^{\pm 2}}{(1+\omega)^{2}} + \sum_{n=2}^{6}  \frac{\alpha^{n}~\Omega^{\pm 2n}}{(1+\omega)^{2n}},
\end{equation}
At early (large Hubble) stage the gravitational inflation powers the cosmic inflation by the quadratic $T^{2}$ correction as appears in the $\alpha$ terms. While the first term of the scalar potential (\ref{lead_pot}) represents a pure kinetic term which is sensitive only to the EoS parameter $\omega$. At this stage the kinetic term becomes more important, the thermal EoS changes lead the universe to deviate from de Sitter universe producing density inhomogeneities at later stages which has an important impact on the galaxy formation.
\subsection{Higher order corrections of the scalar potential}\label{S6.2}
\begin{figure}
\centering
\subfigure[figtopcap][effective potential pattern according to thermal $\omega$ correction]{\label{fig5a}\includegraphics[scale=.3]{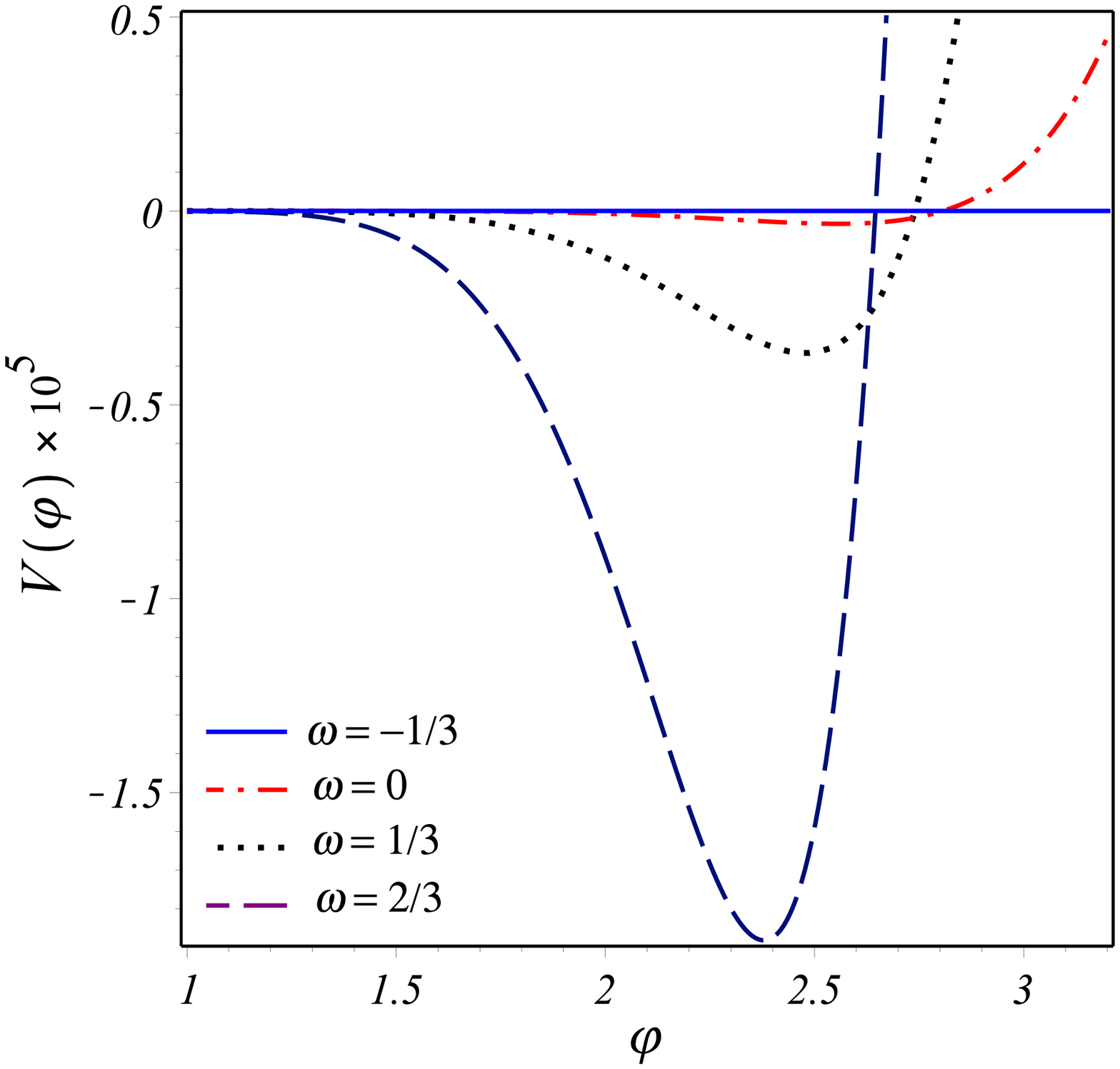}}\hspace{2cm}
\subfigure[figtopcap][effective potential pattern according to $\alpha$ correction]{\label{fig5b}\includegraphics[scale=.3]{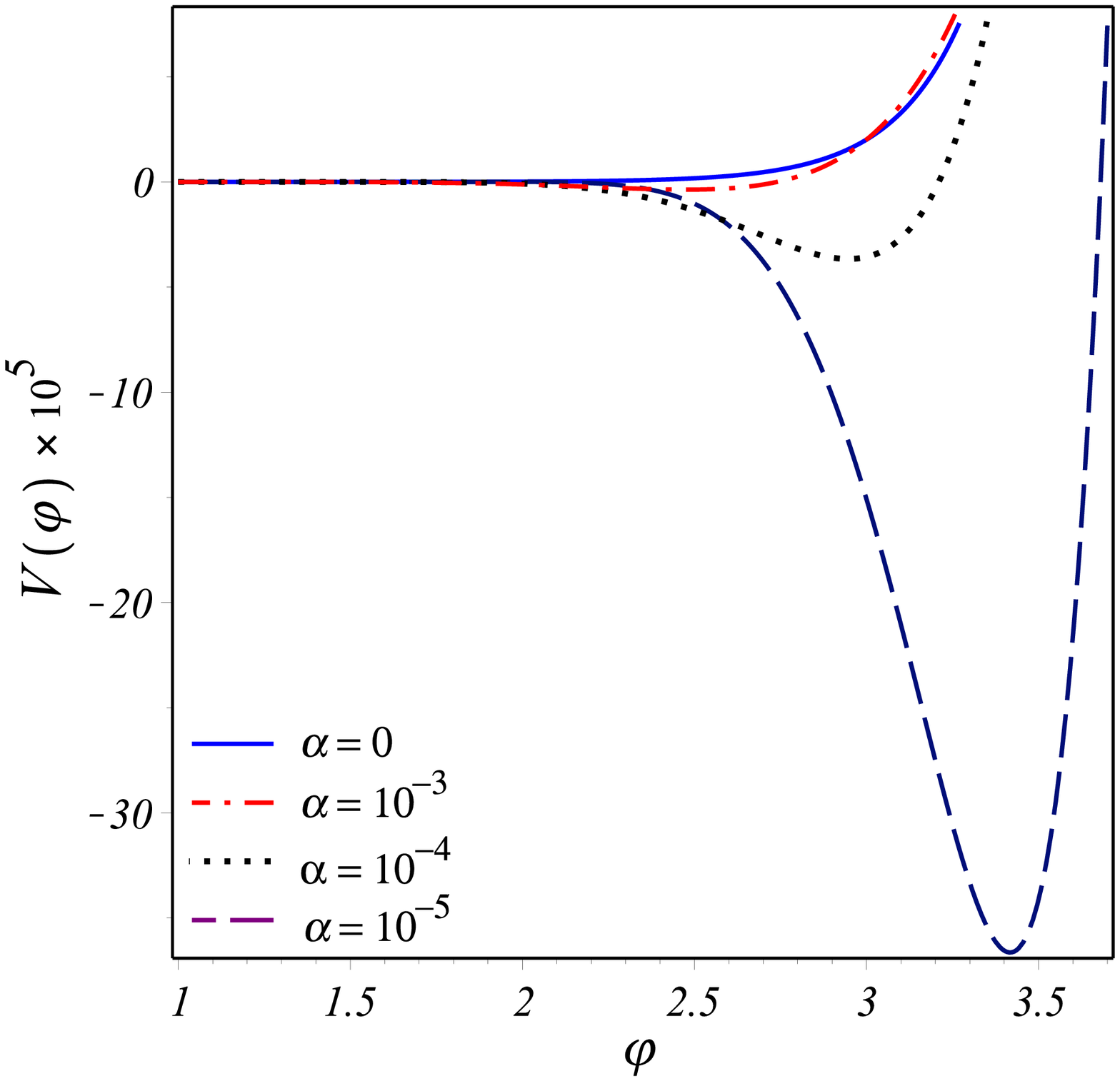}}
\caption[figtopcap]{The correction term of the scalar field (\ref{phi}) contributes as a thermal correction:
\subref{fig5a} the scalar field potential at fixed value of the parameter $\alpha=10^{-3}$ with different values of the EoS parameter $\omega$;
\subref{fig5b} the scalar field potential at fixed EoS parameter $\omega=1/3$ with different values of the parameter $\alpha$.
The initial values have been chosen as $t_{0}=10^{-5}$ and $\varphi_{0}=0$.}
\label{Fig5}
\end{figure}

In the previous Subsection \ref{S6.1} we discussed the contribution of the EoS parameter $\omega$ in the scalar field (\ref{phi}) separately from the parameter $\alpha$. While the parameter $\alpha$ appears as a gravitational pure contribution. In this Subsection we use the full expression of the parameterized time (\ref{phi}) to study possible higher corrections due to the combination of the two parameters $\alpha$ and $\omega$. So the parameterized time due to (\ref{phi}) can be written as\footnote{where the LambertW function is defined as
$$\textmd{\small LambertW}(x)\sim \log(x)-\log[\log(x)]+\sum_{m = 0}^{\infty}\sum_{n = 0}^{\infty}C^{m}_{ n}\frac{\log[\log(x)]^{m+1}}{\log(x)^{m+n+1}}.$$}
\begin{equation}\label{time1}
    t=t_{0}+\underbrace{e^{\textmd{\small LambertW}\left[\frac{8\alpha e^{\frac{3\sqrt{6}(1+\omega)}{2}(\varphi-\varphi_{0})}}{(1+\omega)^{2}}\right]^{\frac{1}{2}}}}_{\textmd{correction term}} \times e^{\pm \frac{3\sqrt{6}(1+\omega)}{4}(\varphi-\varphi_{0})},
\end{equation}
where $\textmd{LambertW}(0)=0$; then the correction term of parameterized time (\ref{time1}) approaches unity at the low energy limit $\alpha \rightarrow 0$; then the parameterized time expression (\ref{time1}) reduces to (\ref{time}) up to the leading term. We perform some calculations similar to the previous Subsection allowing to evaluate the full expression of the effective potential
\begin{eqnarray}\label{Thermalpot}
\nonumber    &&V(\varphi,\alpha,\omega)=\\
\nonumber    &&\frac{4}{27(1+\omega)^{2}}e^{-\frac{1}{2}\sqrt{\frac{2}{3}}X}
    +\alpha\left[\frac{64(\pi-\frac{9}{16}-\frac{9}{8}\omega)}{27\pi(1+\omega)^{4}}e^{-\sqrt{\frac{2}{3}}X}\right.\\
\nonumber    &+&\frac{256\alpha(\pi-\frac{27}{4}-\frac{45}{4}\omega)}{27\pi(1+\omega)^{6}}e^{-\frac{3}{2}\sqrt{\frac{2}{3}}X}
    -\frac{3584\alpha^{2}(\omega+\frac{4}{7})}{3\pi(1+\omega)^{8}}e^{-2\sqrt{\frac{2}{3}}X}\\
    &-&\left.\frac{4096\alpha^{3}(\omega+\frac{1}{3})}{\pi(1+\omega)^{10}}e^{-\frac{5}{2}\sqrt{\frac{2}{3}}X}
    +\frac{16384\alpha^{4}}{3\pi(1+\omega)^{12}}e^{-3\sqrt{\frac{2}{3}}X}\right],
\end{eqnarray}
where $X:=\textmd{LambertW}\left[\frac{8\alpha e^{\frac{3\sqrt{6}}{2}(1+\omega)(\varphi-\varphi_{0})}}{(1+\omega)^2}\right]^{\frac{\sqrt{6}}{9}}-(1+\omega)(\varphi-\varphi_{0})$. It is clear that the effective potential reduces to $V_{l}$($\psi$, $\alpha$, $\omega$) when the correction term in $X$ is negligible. The full expression (\ref{Thermalpot}) allows to evaluate the kinetic term at the high energy limit of the early stage as appears in the first term. This might give some constraints on the initial inhomogeneities of the universe. The sensitivity of the effective potential to its both parameters is shown in Figure \ref{Fig5}. In conclusion, we see that this model is parameterized by two parameters $\omega$ and $\alpha$ both derives the universe to exit the de Sitter universe. The first provides a thermal-like correction, as $V(\varphi_{\textmd{min}}) \geq 0$ as $\omega<0$, it works effectively with the kinetic energy at late stages of the low Hubble regime allowing late phase transitions from radiation to cold dark matter, see Figure \ref{Fig5}\subref{fig5a}. The second parameter $\alpha$ is purely gravitational, it works effectively at large Hubble regime allowing a graceful exit out of de Sitter by reheating of slow roll potential, see Figure \ref{Fig5}\subref{fig5b}.
\section{The slow roll analysis}\label{S7}
Assuming that the inflation epoch is dominated by the scalar field potential only. The slow-roll models defines two parameters as
\begin{equation}\label{slow_roll}
    \epsilon=\frac{1}{16\pi}\left(\frac{V'}{V}\right)^{2},\qquad \eta=\frac{1}{8\pi}\left(\frac{V''}{V}\right).
\end{equation}
These parameters are called slow roll parameters. Consequently, the slow-roll inflation is valid where $\epsilon \ll 1$ and $|\eta| \ll 1$ when the potential is dominating. While the end of inflation is characterized by $\textmd{Max}(\epsilon,|\eta|)=1$ as the kinetic term contribution becomes more effective. The slow roll parameters define two observable parameters
\begin{equation}
\nonumber    r=16\epsilon,\quad n_{s}=1-6\epsilon+2\eta,
\end{equation}
where $r$ and $n_{s}$ are called the tensor-to-scalar ratio and scalar tilt (spectral index), respectively. Recent observations by Planck and BICEP2 measure almost the same scalar tilt parameters $n_{s}\sim 0.96$. However, Planck puts an upper limit to the $r$-parameter $<0.11$ which supports models with small $r$. While BICEP2 sets a lower limit on the $r$-parameter $>0.2$ which supports inflationary models with large $r$. We devote this Section to investigate the capability of the slow roll models to perform both Planck and BICEP2.
\subsection{Reconstruct a potential from Planck and BICEP2}\label{S7.1}
In order to reconstruct a scalar potential performing Planck and BICEP2 data. It is clear that they agree on the spectral index parameter $n_{s}=0.963$ while they give different tensor-to-scalar ratios $r$. We found that, if the slow roll parameters (\ref{slow_roll}) satisfy the proportionality relation $\epsilon=\kappa \eta^{2}$, where $\kappa$ is a constant coefficient. This gives a chance to find two values of $\eta$ perform the same $\epsilon$. Consequently, two values of $r$ for a single value of $n_{s}$
\begin{equation}
\nonumber    r=16\epsilon,\quad n_{s}=1-6(\epsilon=\kappa\eta^{2})+2\eta.
\end{equation}
Interestingly, if the slow roll parameters are constrained by this relation, this produces a simple differential equation with a solution
\begin{equation}
\nonumber    V(\varphi)=A+Be^{-\sqrt{\frac{1}{2\kappa}}\varphi},
\end{equation}
where $A$ and $B$ are constants of integration. In this way, we found that Starobinsky model might be reconstructed naturally from observations if we assume Planck and BICEP2 are correct.
\subsection{The slow roll parameters of the model}\label{S7.2}
\begin{figure}
\centering
\includegraphics[scale=.5]{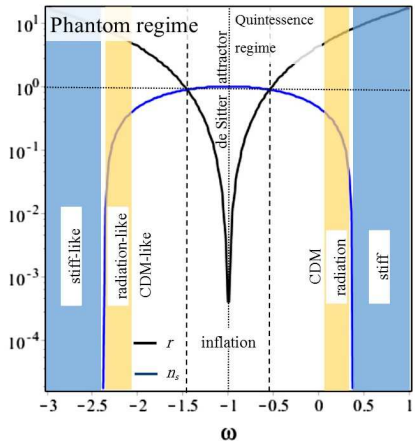}
\caption{The observational inflationary parameters ($r,n_{s}$) according to the thermal EoS correction. Their evolutions are symmetric about the phantom divided line $\omega=-1$ so that it is impossible to distinguish between phantom and non-phantom regimes. The gravitational parameter is chosen as $\alpha=10^{-4}$.}
\label{Fig6}
\end{figure}

We use the standard slow-roll approximation technique to analyze the $T^{2}$ model. Assuming that the inflation epoch is dominated by the scalar field potential only. Using (\ref{sc_pot}) and (\ref{slow_roll}), we evaluate the slow roll parameters
\begin{eqnarray}
  &&\epsilon = \frac{1}{4\pi}\left[\sum_{n=0}^{5}\alpha^{n}\xi_{n}e^{2n\psi}/\sum_{n=0}^{5}\alpha^{n}c_{n}e^{2n\psi}\right]^{2}\sim\\
\nonumber &&\frac{27(1+\omega)^{2}}{32\pi}\left[\frac{\pi(1+\omega)^{2}-18\alpha(2\omega-1)
  e^{-\frac{3\sqrt{6}}{2}(1+\omega)(\varphi-\varphi_{0})}}{\pi(1+\omega)^{2}-9\alpha(2\omega-1)
  e^{-\frac{3\sqrt{6}}{2}(1+\omega)(\varphi-\varphi_{0})}}\right]^{2},\label{slow1}\\[8pt]
  &&\eta = \frac{1}{2\pi}\left[\sum_{n=0}^{5}\alpha^{n}\chi_{n}e^{2n\psi}/\sum_{n=0}^{5}\alpha^{n}c_{n}e^{2n\psi}\right]\sim\\
\nonumber &&\frac{27(1+\omega)^{2}}{16\pi}\left[\frac{\pi(1+\omega)^{2}-36\alpha(2\omega-1)
  e^{-\frac{3\sqrt{6}}{2}(1+\omega)(\varphi-\varphi_{0})}}{\pi(1+\omega)^{2}-9\alpha(2\omega-1)
  e^{-\frac{3\sqrt{6}}{2}(1+\omega)(\varphi-\varphi_{0})}}\right],\label{slow2}
\end{eqnarray}
where $\xi_{n}$, $\chi_{n}$ and $c_{n}$ are known constants. The number $e$-folds from the end of inflation to the time of horizon crossing for observable scales
\begin{eqnarray}\label{efold}
\nonumber    N_{*}(\varphi)&=&-8\pi\int_{\varphi}^{\varphi_{f}}{\frac{V}{V'}}d\varphi\\
\nonumber    &\sim&\frac{8\pi}{9(1+\omega)^{2}}\left[\sqrt{6}(1+\omega)\varphi
    +\frac{1}{3}\ln\left(\pi(1+\omega)^{2}\right.\right.\\
    &-&\left.\left.18\alpha(2\omega-1)e^{-\frac{3\sqrt{6}}{2}(1+\omega)\varphi}\right)\right],
\end{eqnarray}
where $\varphi_{f}$ is the value of $\varphi$ at the end of inflation, i.e. Max($\epsilon$, $|\eta|$)=1. Using (\ref{slow1})-(\ref{efold}) we can reexpress the slow roll parameters as functions of $N$. For the allowed range $30 < N_{*} < 60$, the observable cosmological wavelengths exit the Hubble radius for a minimum and maximum inflationary scale, we take $N_{*}=50$. Next, we calculate the tensor-to-scalar ratio $r=16\epsilon_{*}$ and the scalar tilt $n_{s}=1-6\epsilon_{*}+2\eta_{*}$. The thermal correction is given by the plots of Figure \ref{Fig6}. We see that the $r$ and $n_{s}$ parameters evolve symmetrically towards an attractor at $\omega=-1$. In other words, for each cosmic state $\omega>-1$ there is an identical reflected copy at $\omega<-1$, which is in agreement with the common belief that the phantom ($\omega<-1$) and non-phantom ($\omega>-1$) regimes are indistinguishable. Also, the $r$-plot is highly sensitive to the $\omega$-value. We highlight some important predictions show the capability of the model to perform large tensor-to-scalar parameter as well as small ones.
\begin{table}
\caption{Capability of the model to calculate $r$ and $n_{s}$ parameters}
\label{T1}
\begin{tabular*}{\columnwidth}{@{\extracolsep{\fill}}lccc@{}}
\hline
\multicolumn{1}{c}{$\omega$} & \multicolumn{1}{c}{$r$} & \multicolumn{1}{c}{$n_{s}$} & \multicolumn{1}{c}{Theoretical or}\\
&&&observational \\
\hline
$-1$ & $0$ & $1$ & $\Lambda$ de Sitter \\
$-1\pm 0.03$ & $4.81\times 10^{-3}$ & $0.9994$ & Starobinsky\cite{BO14} \\
$-1\pm 0.16$ & $0.110$ & $0.9862$ & Planck \cite{1303.5082} \\
$-1\pm0.24$ & $0.248$ & $0.9691$ & BICEP2 \cite{1403.3985}\\
\hline
\end{tabular*}
\end{table}
The values in Table \ref{T1} are calculated at $\alpha=10^{-4}$, up to $O(\alpha^{2})$ given by the last expressions in (\ref{slow1}) and (\ref{slow2}). From Table \ref{T1}, we see that the model is sensitive to the choice of the thermal EoS value. If the universe is dominated by $\omega=-1\pm0.16$, this matches perfectly the Planck observations. Nevertheless, the model might support the measured value of $r$ if the universe is dominated by $\omega=-1\pm0.24$ which is unlikely at the present time.\\

In conclusion, one can see that the slow roll parameters (\ref{slow1}) and (\ref{slow2}) are related by $\epsilon \sim \kappa \eta^{2}$. Similar relation has been obtained in the literature when studying the leading term behaviour of the Starobinsky inflation as a special case of the $T$-models \cite{CGP14}. As mentioned in Subsection \ref{S7.1}, the model satisfies the relation, we can write the scalar tilt parameter by a quadratic polynomial in $\eta$ as $n_{s} \sim 1+2\eta-6\kappa\eta^{2}$. There is an agreement on the value of the parameter $n_{s}\sim 0.963$, so the model perform two values of the slow roll parameter $\eta$ holding the same value of the scalar tilt parameter. Consequently, these models perform two different values of the second slow roll parameter $\epsilon$ so the tensor-to-scalar ratio $r=16\epsilon$. Recalling Figure \ref{Fig4}\subref{fig4a}, the false vacuum is separated by a broad barrier. However, the top of the barrier is quite flat.the decay of the false vacuum is followed by slow-roll inflation allowing a tunneling event from the high energy false vacuum. Actually, this pattern suggested to perform both Planck and BICEP2 data \cite{BHS13}. When a negative value of $\eta$ is observed near the peak of $\varphi$, it would need to be offset by a positive value of $\eta$ at some later time over a comparable field range in order to get $\epsilon$ to be small again during the period of observable inflation. For example, when use the value $n_{s}\sim 0.963$ as an input for a particular choice of $\kappa \sim 30$. So we find
\begin{itemize}
\item [(i)] The first has a negative value of $\eta \sim -9.34\times10^{-3}$ which leads to a small $\epsilon \sim 3.05\times10^{-3}$.
\item [(ii)] The second solution of $\eta$ has a positive value of $\sim 2.09\times10^{-2}$ which gives $\epsilon \sim 1.31\times10^{-2}$.
\end{itemize}
Surely both negative and positive values of $\eta$ give the same spectral index $n_{s} \sim 0.963$. Nevertheless, we can get two simultaneous tensor-to-scalar ratios: the first is too small $r \sim 4.9\times10^{-2}$, while the other value is larger $\sim 0.21$. We conclude that the slow roll inflationary models which are characterized by the proportionality $\epsilon \propto \eta^{2}$ can perform both Planck and BICEP2.
\section{Summary}\label{S8}
We have applied a quadratic $f(T)$ field equations to the flat FRW universe with a perfect fluid. The choice of $f(T)=T+\alpha T^{2}$ has enabled us to modify the Friedmann equations, whereas the density and pressure of two fluids of matter and torsion reproduce the GR as $\alpha=0$. We have assumed the matter is governed by the continuity equation. We have studied two cases of the matter choices:
\begin{itemize}
\item [(i)] The universe is assumed to be dominant by $\Lambda$DE with EoS $\omega=-1$. The solution provided an exponential scale factor $a_{\textmd{inf}}(t)=a_{0}e^{H_{0}(t-t_{0})}$, which fixes the torsion EoS to a value of $\omega_{T}=-1$. The standard cosmological study has given Hubble and deceleration parameters as $H=const.$ and $q=-1$. So this model is consistent with $\Lambda$ de Sitter model. Also, we found that the choice of $\omega=-1$ implying that $f(T) \sim T \sim \Lambda$. Consequently, the coefficient $\alpha$ has been estimated as $\alpha=\frac{1}{12 \Lambda} \sim 10^{-73}~s^{2}$. So the model omitted the quadratic term $T^{2}$ anyways. We found that the model can predict an eternal inflationary scenario.
\item [(ii)] The universe is assumed to have an EoS $\omega \neq -1$. The solution provided a decomposable effective scale factor, $a_{\textmd{eff}}(t) \propto a_{\textmd{v}} a_{m}$, where $a_{\textmd{v}}$ having an exponential form of time allows an inflation period limited to a specific time interval, whereas $a_{m} \propto t^{\frac{2}{3(1+\omega)}}$ matches perfectly the matter dominant universe epoch. So the effective scale factor allows a graceful exit inflation model. However, the model does not assume slow roll conditions. The extensive analysis of the scale factor indicates that the vacuum ends the inflation phase then enters an endless phase transition releasing its latent heat. By the end of this stage the effective scale factor matches perfectly the matter scale factor announcing a matter dominant universe to showup.\\
\end{itemize}

The cosmological study of the quadratic $f(T)$ theory shows that at the large Hubble regime of the early universe is expanding with an acceleration as $H>0$ and $q<0$ which matches the cosmic inflationary models. Nevertheless, at the low Hubble regime of the late universe matches the decelerated FRW as $H>0$ and $q>0$. On the other hand, the effective EoS parameter evolves as $\omega_{\textmd{eff}}:~ \omega_{T}=-1 \rightarrow \omega$, the transition from early negative EoS to positive value of the ordinary matter assures the transition from an accelerated phase to later decelerated FRW phase.\\

The quadratic torsion $T^{2}$ induces a pure gravitationally cosmic inflation. However, the torsion scalar shows a decaying behaviour. This shares some features with the scalar field theory of inflation. Unlike $f(R)$ theories, the $f(T)$ theories cannot be conformally transformed to TEGR plus scalar field. So it is not expected that the quadratic $f(T)=T+\alpha T^{2}$ to be equivalent to $f(R)=R+\alpha R^{2}$ which is known as the Starobinsky model. We applied a special treatment by constructing the torsion tensor from a scalar field $\varphi$. This treatment shows a kinetic coupling between the scalar field and spacetime torsion. On another word, we relate the scalar field to the vierbein field, i.e. the space-time symmetry. Moreover, the treatment allows to map the torsion density and pressure to scalar field model without conformal transformation. In this way, we able to construct an effective potential of the scalar field from the adopted $f(T)$ gravity theory. Finally, we reformulate the inflationary background in terms of this scalar field $\varphi$ in a potential $V(\varphi)$. The effective potential derives the early inflation epoch, while the kinetic coupling between the scalar field and spacetime torsion explains the rushing the effective EoS from $\Lambda$DE to radiation dominant universe. So the results are consistent with the effective scale factor analysis.\\

The evaluated potential shows a tunneling event after the false vacuum decay followed by a slowly rolling pattern to an effective minimum of the true vacuum. These types of potentials are suggested to perform Planck and BICEP2 results. The model is characterized by two parameters ($\omega,\alpha$): The first contributes to the kinetic term at late stages as a thermal correction, the second contributes to the effective potential as a purely gravitational effect at early stages. Both parameters derive the universe to exit the de Sitter inflation through reheating process at the effective minimum potential which agrees with the scale factor analysis.\\

Finally, we show that the slow roll models which are characterized by the proportionality $\epsilon \propto \eta^{2}$ can perform Planck and BICEP2 results. The evaluated slow roll parameters show that the model fulfills the above mentioned proportionality. The evolution of the observable parameters ($r,n_{s}$), according to the thermal EoS changes, produces de Sitter as an attractor with ($r,n_{s}$) $=$ ($0,1$) at the phantom divided line $\omega=-1$. The model satisfies Planck data if the thermal state is dominated by $\omega\sim-1\pm0.16$. However, it satisfies BICEP2 data if the cosmic thermal state is dominated by $\omega\sim-1\pm0.24$ which is less likely. Also, the parameters evolve symmetrically about the de Sitter attractor, producing an indistinguishable phantom ($\omega<-1$) and non-phantom ($\omega>-1$) regimes.
\subsection*{acknowledgments}
This work is partially supported by the Egyptian Ministry of Scientific Research under project No. 24-2-12.
\bibliographystyle{aiaa}

\end{document}